\documentclass[aps,prd,twocolumn,preprintnumbers,superscriptaddress,dblfloatfix,nofootinbib]{revtex4-2}
\usepackage[utf8]{inputenc}
\usepackage{lmodern}
\usepackage{amsmath,amssymb,mhchem}
\usepackage{graphicx}
\usepackage{url}
\usepackage{color}
\usepackage{enumitem}
\usepackage{subfigure}
\usepackage[dvipsnames]{xcolor}
\usepackage[colorlinks=true,breaklinks=true]{hyperref}
\hypersetup{allcolors=[rgb]{0.0 0.0 0.6},linkcolor=[rgb]{0.75 0.05 0.05}}
\usepackage{bm}
\usepackage{epsfig}
\usepackage{amsmath}
\usepackage{amssymb}
\usepackage{slashed}
\usepackage{color}
\usepackage{accents}
\usepackage[dvipsnames]{xcolor}
\usepackage[colorlinks=true,breaklinks=true]{hyperref}
\hypersetup{allcolors=[rgb]{0.0 0.0 0.6},linkcolor=[rgb]{0.75 0.05 0.05}}
\usepackage[export]{adjustbox}

\newcommand{\Msun}{M_{\odot}}
\newcommand{\beq}{\begin{equation}}
\newcommand{\eeq}{\end{equation}}
\def\bea{\begin{eqnarray}}
\def\eea{\end{eqnarray}}
\newcommand{\bei}{\begin{itemize}}
\newcommand{\eei}{\end{itemize}}
\newcommand{\bmat}{\begin{matrix}}
\newcommand{\emat}{\end{matrix}}

\newcommand{\Fig}[1]{Fig.~\ref{#1}}
\newcommand{\Eq}[1]{Eq.~(\ref{#1})}

\def\={\,=\,}
\def\+{\,+\,}
\def\-{\,-\,}

\begin{document}

\preprint{KEK-QUP-2023-0032, KEK-TH-2579, KEK-Cosmo-0333, IPMU23-0045}

\title{Coexistence Test of Primordial Black Holes and Particle Dark Matter\\ from Diffractive Lensing}

\author{Han Gil Choi}
\email{hgchoi1w@gmail.com}
\affiliation{Cosmology, Gravity and Astroparticle Physics Group, Center for Theoretical Physics of the Universe,
Institute for Basic Science (IBS), Daejeon, 34126, Korea}

\author{Sunghoon Jung}
\email{sunghoonj@snu.ac.kr}
\affiliation{Center for Theoretical Physics, Department of Physics and Astronomy, Seoul National University, Seoul 08826, Korea}
\affiliation{
Astronomy Research Center, Seoul National University, Seoul 08826, Korea
}

\author{Philip Lu}
\email{philiplu11@gmail.com}
\affiliation{Center for Theoretical Physics, Department of Physics and Astronomy, Seoul National University, Seoul 08826, Korea}

\author{Volodymyr Takhistov}
\email{vtakhist@post.kek.jp}
\affiliation{International Center for Quantum-field Measurement Systems for Studies of the Universe and Particles (QUP, WPI),
High Energy Accelerator Research Organization (KEK), Oho 1-1, Tsukuba, Ibaraki 305-0801, Japan}
\affiliation{Theory Center, Institute of Particle and Nuclear Studies (IPNS), High Energy Accelerator Research Organization (KEK), Tsukuba 305-0801, Japan
}
\affiliation{Graduate University for Advanced Studies (SOKENDAI), \\
1-1 Oho, Tsukuba, Ibaraki 305-0801, Japan}
\affiliation{Kavli Institute for the Physics and Mathematics of the Universe (WPI), UTIAS, \\The University of Tokyo, Kashiwa, Chiba 277-8583, Japan}
 
\date{\today}

\begin{abstract}
If dark matter (DM) consists of primordial black holes (PBHs) and particles simultaneously, PBHs are generically embedded within particle DM halos. 
Such ``dressed PBHs'' (dPBHs) are subject to modified constraints compared to PBHs and can contribute to significant DM abundance in the mass range $10^{-1} \textendash 10^2 \Msun$. We show that diffractive lensing of chirping gravitational waves (GWs) from binary mergers can not only discover, but can also identify dPBH lenses and discriminate them from bare PBHs on the event-by-event basis, with potential to definitively establish the coexistence of subdominant PBHs and particle DM.
\end{abstract}

\maketitle

{\it Introduction}\textemdash 
Primordial black holes (PBHs) formed in the early Universe constitute
a compelling dark matter (DM) candidate~(e.g.~\cite{Zeldovich:1967,Hawking:1971ei,Carr:1974nx,Meszaros:1975ef,Carr:1975qj,GarciaBellido:1996qt,Kawasaki:1997ju,Kohri:2007qn,Khlopov:2008qy,Frampton:2010sw,Bird:2016dcv,Kawasaki:2016pql,Inomata:2016rbd,Pi:2017gih,Garcia-Bellido:2017aan,Georg:2017mqk,Kocsis:2017yty,Ando:2017veq,Cotner:2016cvr,Cotner:2019ykd,Cotner:2018vug,Sasaki:2018dmp,Carr:2018rid,Kusenko:2020pcg,Carr:2020gox,Green:2020jor,Escriva:2022duf,Kawana:2021tde,Lu:2022jnp,Kawana:2022lba,Lu:2022paj,Lu:2022yuc,Kawana:2022olo,Chakraborty:2022mwu}). The mergers of stellar-mass $\sim 10 \textendash 10^2 \Msun$ PBHs have been linked to the recently discovered gravitational wave (GW) events by LIGO-VIRGO-Kagra (LVK)~\cite{Bird:2016dcv,Sasaki:2018dmp}. These PBHs can contribute to a substantial fraction of DM mass density $f_{\rm PBH} = \Omega_{\rm PBH}/\Omega_{\rm DM}$ (e.g.~\cite{Kohri:2018qtx,Ali-Haimoud:2016mbv,Zumalacarregui:2017qqd,Koushiappas:2017chw,Inoue:2017csr,Manshanden:2018tze,Lu:2020bmd,Serpico:2020ehh,Takhistov:2021aqx,Takhistov:2021upb,Manshanden:2018tze}), with current LVK observations implying $f_{\rm PBH} \lesssim \mathcal{O}(10^{-3})$ assuming PBH mergers~\cite{Bird:2016dcv,Clesse:2016vqa,Sasaki:2016jop,Vaskonen:2019jpv,Franciolini:2021tla}. The origin of merger events is yet to be definitively established. PBHs in the (sub)solar mass range are also expected to produce a variety of intriguing signals, e.g. generated via the interplay of new physics and astrophysics, which can shed light on their origin~\cite{Fuller:2017uyd,Takhistov:2017bpt,Takhistov:2017nmt,Bramante:2017ulk,Takhistov:2020vxs,Dasgupta:2020mqg,Wang:2021iwp,Sasaki:2021iuc,LIGOScientific:2022hai}.

If PBHs compose a subdominant component of the DM abundance, and are embedded in a dominant background of particle DM, a population of PBHs will generically accrete particle DM halos from their surroundings, growing from the recombination era until cosmic structure formation resulting in ``dressed PBHs'' (dPBHs) with a characteristic DM mass profile~\cite{Bertschinger:1985pd,Mack:2006gz,Ricotti:2007au}. 
Note that intermediate-mass BHs can also form distinct particle DM halos at late epochs, also called ``DM spikes"~\cite{Zhao:2005zr,Bertone:2005xz,Bringmann:2009,Aschersleben:2024xsb,Bertone:2024wbn}.
dPBHs constitute a generic signature of primordial origin as well as the coexistence scenario with particle DM. 
dPBHs can be constrained by a variety of model-dependent signals if halo particle DM has significant interactions with the Standard Model~\cite{Adamek:2019gns, Lacki:2010zf,Hertzberg:2020kpm,Nurmi:2021xds}.

There have been several proposals for probing DM halo structures around BHs independent of specific particle DM interactions. Recently, geometrical-optics gravitational lensing of fast radio bursts (FRBs) at cosmological distances was suggested as a promising approach to identify dPBHs~\cite{Oguri:2022fir}. 
The DM halo can also produce significant effects on microlensing constraints~\cite{Cai:2022kbp}, and in the case of extreme mass ratio inspirals, the DM halo around the primary black hole can alter the inspiral motion of the secondary~\cite{Eda:2014kra,Kavanagh:2020cfn,Coogan:2021uqv,Cole:2022ucw,Cole:2022yzw}. However, a general method to distinguish dPBHs from more massive BHs on an individual object basis has not yet been established.

\begin{figure}[ht]
\includegraphics[width=0.7\linewidth]{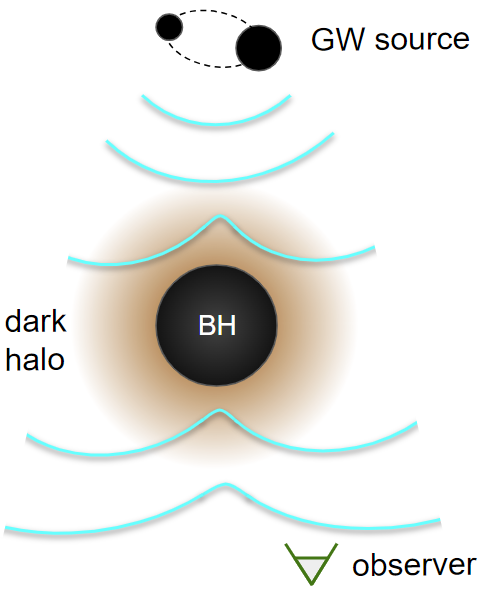}
\caption{ \label{fig:lensing} 
Illustration of the diffractive lensing of chirping GWs from binary mergers by dPBHs at cosmological distances. 
Diffractive lensing distorts GW wave fronts along a broad region of space, which is summed up to produce frequency-dependent amplification of the wave in contrast to the discrete images resulting from geometrical-optics lensing. 
}
\end{figure}

In this Letter, we advance \emph{diffractive} lensing of chirping GWs as a novel method that fully exploits the opportunities to discriminate dPBHs and, thus, enables discovering and establishing their existence (see Fig.~\ref{fig:lensing}). Chirping GWs produced from binary mergers have particularly distinctive waveforms in time and frequency domains, allowing their detection at experiments such as LIGO. 
Chirping GWs were previously demonstrated to be useful in probing small-length-scale objects such as black holes~\cite{Nakamura:1997sw,takahashi2003wave,Jung:2017flg,lai2018discovering,Urrutia:2021qak,Basak:2021ten,LIGOScientific:2021izm,LIGOScientific:2023bwz,Zhou:2022yeo} as well as diffuse dark matter halos~\cite{dai2018detecting,oguri2020probing,Choi:2021bkx,tambalo2022gravitational} by gravitational lensing.
Furthermore, unlike photons, GWs typically experience diffractive lensing, whose nontrivial frequency dependence combined with characteristic waveforms makes it possible to ``scan'' and measure the mass profile of the lensing object directly on an event-by-event basis~\cite{Choi:2021bkx}. 
We demonstrate that with ground-based detectors, GW lensing not only allows the discovery of dPBHs in the mass window of $\sim 10^{-1} \textendash 10^{2}\Msun$, but can also efficiently discriminate them from bare PBHs due to distinctive mass profiles, hence supporting their primordial origin as well as the coexistence scenario with particle DM.

%


{\it Dark halos around black holes}\textemdash  
PBHs that are embedded in a smooth background of particle DM in an expanding Universe will seed the formation of massive DM halos. The halo growth follows the theory of spherical gravitational collapse~\cite{Bertschinger:1985pd}, with infalling DM particles expected to carry angular momentum and not just be radially incorporated into BHs. 
During the radiation-dominated era, the halo mass increases on the order of BH mass $M_h / M_{\rm PBH} \simeq 1$, while during the matter-dominated era the halo mass increases following the cosmological expansion $\propto a = (1 + z)^{-1}$, until the cutoff redshift $z_c \sim 30$, around which large-scale structure forms, affecting the accretion of additional DM and further halo evolution.

The resulting halo mass is~\cite{Mack:2006gz,Berezinsky:2013fxa,Adamek:2019gns,Boudaud:2021irr} 
\begin{equation}
\label{eq:Mh}
    M_h \= 97 \left(\dfrac{31}{1+z_c}\right) M_{\rm PBH}~,
\end{equation}
enclosed within
$R_h =  0.61\textrm{ pc} \left(31/(1+z_c)\right) \left(M_h/M_\odot\right)^{1/3}$.
The halo density profile is given by~\cite{Bertschinger:1985pd,Berezinsky:2013fxa,Boudaud:2021irr} as
 $   \rho_h(r) = \rho_0 \left(R_h/r\right)^{9/4} = 0.26 M_\odot \textrm{ pc}^{-3} \left((1+z_c)/31\right)^3 \left(R_h/r\right)^{9/4}$,
and confirmed by $N$-body simulations~\cite{Serpico:2020ehh,Adamek:2019gns,Boudaud:2021irr}.
$M_h \gg M_{\rm PBH}$ sets the upper limit of the PBH abundance $f_{\rm PBH} \leq M_{\rm PBH}/M_h$ to avoid the overclosure of the Universe.

For lensing, the relevant quantities are the two-dimensional (2D) density profile, line of sight projected onto the lens plane. The total 2D density $\Sigma(\bm{x}) = \Sigma_h + \Sigma_{\rm PBH}$ is given by the DM halo part~\cite{Oguri:2022fir}
\begin{equation}\label{eq:Sh}
    \Sigma_h( \bm{x}) = 2\int_0^{\infty} dz \rho_h(\sqrt{\bm{x}^2+z^2}) \\
    \simeq \rho_0 R_h \sqrt{\pi} \frac{\Gamma(5/8)}{\Gamma(9/8)} \left(\frac{R_h}{|\bm{x}|}\right)^{5/4},
\end{equation}
and the PBH part
\begin{equation}\label{eq:Spbh}
    \Sigma_{\rm PBH}(\bm{x}) \,\simeq\, \frac{M_{\rm PBH}}{2\pi |\bm{x}|} \delta (|\bm{x}|).
\end{equation}
The PBH part contributes as a delta-function distribution since the Schwarzschild radius is much smaller than the other length scales. Equivalently, the 2D potentials, obeying the 2D Poisson equation $\nabla^2 \psi(\bm{x}) =2 \Sigma(\bm{x})/\Sigma_\text{crit}$, are
    $ \psi_h(\bm{x}) \= (32/9)  (\rho_0 R_h^3/\Sigma_\text{crit}) \pi^{1/2}(\Gamma(5/8)/\Gamma(9/8))\left( |\bm{x}|/R_h\right)^{3/4}$
for the DM halo and 
    $ \psi_\text{PBH}(\bm{x}) \= (M_\text{PBH}/\pi \Sigma_\text{crit})\ln |\bm{x}| $
for the PBH. 
Here, $\Sigma_\text{crit} = (4\pi G d_\text{eff})^{-1} $ is the lensing critical density, with $G$ being the gravitational constant and $d_\text{eff} = d_l d_{ls}/d_s$ being the effective lens distance where $d_l$, $d_s$, and $d_{ls}$ are the angular-diameter distances to the lens, the source, and between the lens and source, respectively. 

We assume both sources and lenses are at cosmological distances, where the DM halo significantly affects GW lensing. 
The DM halo contribution to lensing effects can be estimated by Einstein radius $x_E$ of dPBHs. As only the lens mass enclosed within the Einstein radius is relevant to lensing, $x_E$ of a dPBH should be much larger than that of a bare PBH $x_E^{\rm PBH}=\sqrt{4GM_{\rm PBH} d_{\rm eff}} = 0.014\, \mathrm{pc}(M_\mathrm{PBH}/M_\odot)^{1/2}(d_\mathrm{eff}/\mathrm{Gpc})^{1/2}$when the DM halo dominates the GW lensing signal. The Einstein radius of dPBH is approximately
\begin{equation}
    x_E \,\simeq\, x_E^{\mathrm{PBH}}\left[1+4.1 \left(\frac{M_\mathrm{PBH}}{M_\odot}\right)^{1/5} \left(\frac{d_\mathrm{eff}}{\mathrm{Gpc}} \right)^{3/5}\right]^{1/2}.
\end{equation}
Hence, although $M_h \gg M_{\rm PBH}$, the influence of the DM halo on dPBH lensing becomes apparent only for sources and lenses at cosmological distances. Microlensing of stars in our neighborhood cannot efficiently be used to detect the halos as emphasized in Ref.~\cite{Oguri:2022fir}, while the majority of cosmological GW and FRB lensing events can.

\begin{figure}[t]
\includegraphics[width=0.99\linewidth,valign=t]{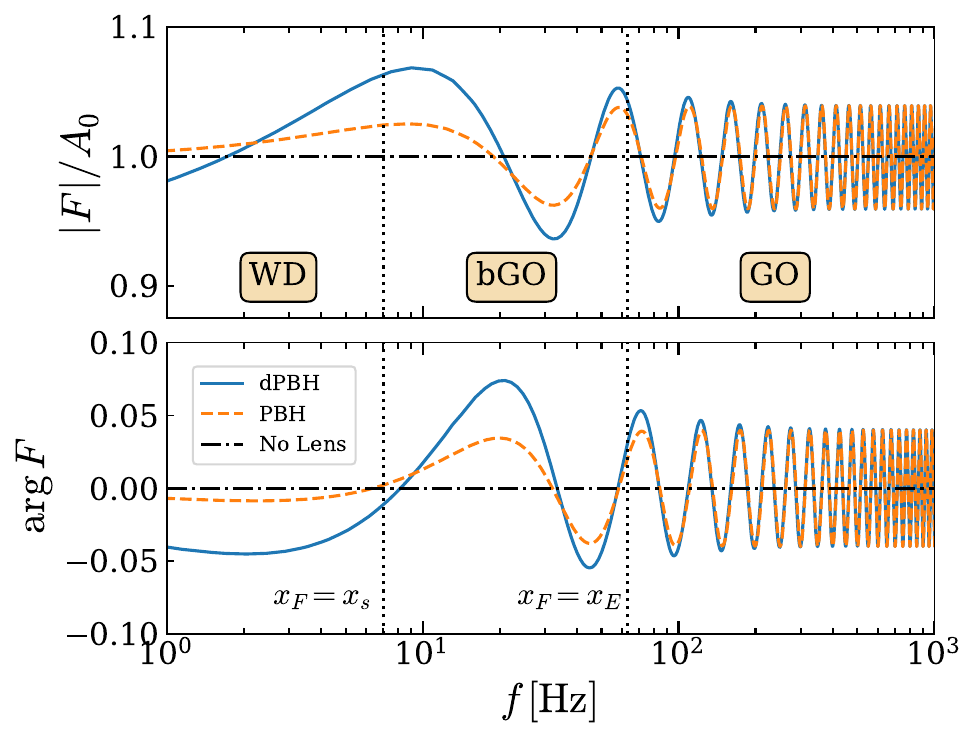}  
\caption{ \label{fig:diffraction}
Frequency dependence of the magnification and phase shift for typical lensing parameters.
Characteristic amplification by dPBH (solid lines) exhibiting nontrivial frequency dependence at low-frequency diffraction regimes and illustrating discrimination from a bare PBH (dashed lines) as well as unlensed observations (straight dot-dashed lines). We assume $M_{\rm PBH}=20\Msun, z_s=0.4$, and $z_l=0.17$. The bare PBH signal is chosen to fit the highest-frequency GO regime, highlighting that low-frequency regimes cannot be simultaneously matched. Regimes for the weak diffraction ($x_F > x_s$), bGO ($x_s > x_F > x_E$), and GO ($x_E > x_F$), that occur for $x_s > x_E$.
For discussion of the case with $x_s < x_E$ see Supplemental Material.
}
\end{figure}

{\it Diffractive lensing}\textemdash 
Lensing is generally quantified by the complex amplification factor $F(f)\equiv h_L(f)/h_0(f)$, defined as the ratio of the lensed and unlensed waveforms in the frequency domain. It is given by the Fresnel-Kirchhoff integral on the 2D lens plane~\cite{deguchi1986wave,nakamura1999wave,takahashi2003wave},
\begin{equation} \label{eq:Fresnel} 
F(f) \= \frac{1}{2\pi i}\int \frac{d^2 \bm{x}}{x_F^2}e^{i \phi(\bm{x},\bm{x}_s) /x_F^2}\, ,
\end{equation}
where $\bm{x}$ is the physical coordinate on the lens plane, with $\bm{x}_s$ being the projected source direction. The characteristic length scale of the lens system is the Fresnel length~\cite{Macquart:2004sh,oguri2020probing,Choi:2021bkx}
\beq
x_F = \sqrt{\frac{d_\text{eff}}{2\pi f(1+z_l)}} =  0.39\, \mathrm{pc} \left(\frac{d_\mathrm{eff}}{\mathrm{Gpc}}\right)^{1/2}\left(\frac{10\, \mathrm{Hz}}{f(1+z_l)}\right)^{1/2}.
\eeq
This is essentially a path integral, superposing all waves passing and bending at points on the lens plane, 
weighted by the propagation phase 
$    \phi(\bm{x},\bm{x}_s) = \frac{1}{2}|\bm{x}-\bm{x}_s|^2 - \psi(\bm{x}) -\phi_m(\bm{x}_s)\, .$
The phase has stationary points, balanced by the positive geometric distance and the negative gravitational Shapiro time delay $\psi(\bm{x})$; a constant offset $\phi_m(\bm{x}_s)$ is included to set the minimum of the phase zero. When the path integral is dominated by discrete stationary points in the high-frequency regime, the result is the usual geometrical-optics (GO) lensing producing discrete images. For weak gravity, this limit is given by $x_F \lesssim x_s$.

On the other extreme of low frequencies with $x_F \gtrsim x_s$, the integral receives contributions from finite regions around each stationary point or from a much broader region. Hence, resulting images blur or overlap. The size of the support is roughly given by $x_F \propto f^{-1/2}$~\cite{oguri2020probing,Choi:2021bkx}, resulting in frequency-dependent amplification. This frequency-dependent regime corresponds to wave optics or diffractive lensing. We note that $x_F$ of the LIGO-band GW-dPBH lensing system is comparable to the $x_E$ of the lens profile. This quantity is of the order of ${\cal O}(0.1\textendash 1)$ pc. Thus, diffractive lensing is relevant to our study.

An example of frequency dependence of the magnification and phase shift for typical lensing parameters in our study is displayed in \Fig{fig:diffraction}.
The nontrivial modulation of the typical GW lensing signal magnification with respect to the unlensed value $F/A_0=1$ is larger than $\sim {\cal O}(10)\%$, readily detectable at ground-based detectors with chirping GWs (see Supplemental Material). 

{\it Halo discrimination}\textemdash  
We now discuss how diffractive lensing of chirping GWs measures and discriminates lens mass profiles.
Diffractive lensing is generally relevant for $x_F \gtrsim {\cal O}(x_s, x_E)$. 
In the \emph{weak diffraction} regime $x_F \gtrsim x_s$ (here,  $x_s > x_E$; the other case with $x_s < x_E$ will be commented on later),
$F(f)$ is approximated by the analytic continuation of $\overline{\kappa}(x)$~\cite{Choi:2021bkx}:
\begin{equation}\label{eq:WD}
    F(f) - 1 \, \simeq \, \overline{\kappa}(x_F e^{i\pi/4})~,
\end{equation}
where $\overline{\kappa}(x)= 2\pi(\pi x^2)^{-1}\int_0^{x} dx' x' \Sigma(x')/\Sigma_\text{crit}$ is the mean convergence (enclosed mass density normalized to the critical value) within an aperture of radius $x$ centered at the lens position.
This remarkable result, evaluating the density profile at the frequency-dependent location (ignoring the constant phase $e^{i \pi/4}$ for order-of-magnitude estimation), clearly shows that the growing (chirping) GW frequencies probe the density profile at successively smaller length scales~\cite{Choi:2021bkx,Jung:2022tzn}.

The general idea and strategy for discriminating dPBHs from bare PBHs are illustrated in \Fig{fig:diffraction}. First, the dPBH data in the highest-frequency GO regime are fit by hypothetical bare PBH lensing with a choice of the bare PBH mass $M_{\rm eff}$ and the impact parameter. This is always possible, since there will be two independent observables in the GO regime of chirping GWs -- the amplitude and frequency of the interference fringe as shown in \Fig{fig:diffraction} or, equivalently, the time-delay and magnification ratio of (interfering) two images. The exact relationship between these two sets of variables is derived in Supplemental Material. There remains an uncertainty on the overall GW amplitude, denoted schematically by $A_0$; thus, \Fig{fig:diffraction} is normalized to oscillate around unity.

After the GW lensing data are matched in the GO regime, the dPBH data in lower-frequency weak diffraction regimes ($x_F> x_s$; the leftmost region in \Fig{fig:diffraction}) cannot be matched simultaneously. 
Weak diffraction effects depend nontrivially on the frequency via the spatially varying enclosed mass of the dPBH halo [see \Eq{eq:WD}], while the enclosed mass of the bare PBH does not change. Thus, the mismatch in the lower-frequency weak diffraction regime (as shown in \Fig{fig:diffraction}) is a distinctive signature of dPBHs in contrast to bare PBHs.

This can be seen analytically. Weak diffraction by the dPBH lens follows
$F(f)-1 \simeq \overline{\kappa}(x_F e^{i\pi/4})\propto f^{5/8}$, from the density profile $\Sigma(x) \propto x^{-5/4}$ in \Eq{eq:Sh}. On the other hand, the bare PBH lens with $\overline{\kappa} \propto x^{-2}$ yields a distinct frequency dependence, $F(f)-1 \propto  f$. 

Furthermore, for the cases with $x_s > x_E$, the beyond-GO regime (bGO), the transition between weak diffraction and GO for $x_s>x_F>x_E$, can also contribute significantly. This is where the quadratic approximation near saddle points is not accurate enough~\cite{takahashi2004quasi,tambalo2022gravitational,savastano2023weakly}. Higher derivatives of the mass profile around the primary image yield bGO corrections to $\delta F(f)  \propto C f^{-1}$, 
with $C$ capturing the higher derivatives at the image.
Thus, distinct mass profiles result in different 
bGO corrections as shown in \Fig{fig:diffraction}
(see also Supplemental Material for the $x_E > x_F > \sqrt{x_E x_s}$ case, i.e. the strong diffraction regime).

 In all, discrimination is expected to be possible if both the low-frequency weak diffraction or bGO and the high-frequency GO regimes of the GW lensing event are well measured. The different behaviors of these lensing regimes are illustrated in \Fig{fig:diffraction} and in Supplemental Material. The transition frequencies between these regimes depend on $x_s,~x_E,$ and $d_{\rm eff}$. For stellar-mass PBHs, the GO regime typically corresponds to $f \gtrsim 100$ Hz, which can be well measured at ground-based detectors, 
although they can potentially probe lower-frequency diffraction regimes as well. Below, we estimate the fraction of lensing events that allow dPBH discrimination with ground-based detectors.

{\it Detection prospects}\textemdash 
GWs from binary mergers have specific evolutions of amplitude and phase, called \emph{chirping}, which show specific frequency dependencies $h_0(f) \propto f^{-7/6} e^{i\Psi(f)}$ and $\Psi(f) \propto f^{-5/3}$~\cite{cutler1994gravitational}.
These predictions provide a valuable basis for detecting tiny nonstandard frequency-dependent effects added to the chirping. For example, diffractive lensing shown in \Fig{fig:diffraction} induces characteristic frequency dependencies, known to be well detected with chirping GWs robust against other effects in binary mergers (such as masses and distances) and higher-order relativistic corrections. 

Lensing detection significance can be conveniently characterized by $\langle \ln \Lambda \rangle$, the expectation value of the log-likelihood ratio between the lensing hypothesis and the no-lensing hypothesis, optimized for constant amplitude rescaling, constant phase shift, and time shift~\cite{LIGOScientific:2019hgc,romano2017detection,Jung:2017flg,dai2018detecting,Choi:2021bkx}. This significance measure incorporates the sensitivity of the GW detector and accounts for the degeneracy between lensing parameters and GW source parameters. Such estimates have been shown to yield accurate results~\cite{Jung:2017flg,dai2018detecting,Choi:2021bkx}.

We use the threshold $\langle \ln \Lambda \rangle>3$ for lensing event detections to estimate the detectable lensing probability for the given GW source~\cite{Jung:2017flg,Choi:2021bkx}. Choosing a different $\langle \ln \Lambda \rangle$ threshold does not significantly change final results~\cite{Jung:2017flg}. The comoving number density of dPBH lenses, $n_{\rm PBH} = f_{\rm PBH} \Omega_{\rm DM} \rho_c/M_{\rm PBH}$, is assumed to be constant. GW sources are assumed to be uniformly distributed, with the BH merger mass function taken from the Power-law and Peak model in~\cite{talbot2018measuring}, normalized by the overall number density $R_0 =28.3\,\text{Gpc}^{-3}\text{yr}^{-1}$ suggested by GW observations~\cite{abbott2019binary,abbott2021population,KAGRA:2021duu}. We consider GW sources up to $z_s \leq 10$  (see Supplemental Material for details).

\begin{figure}[t] 
\includegraphics[width=0.99\linewidth,valign=t]{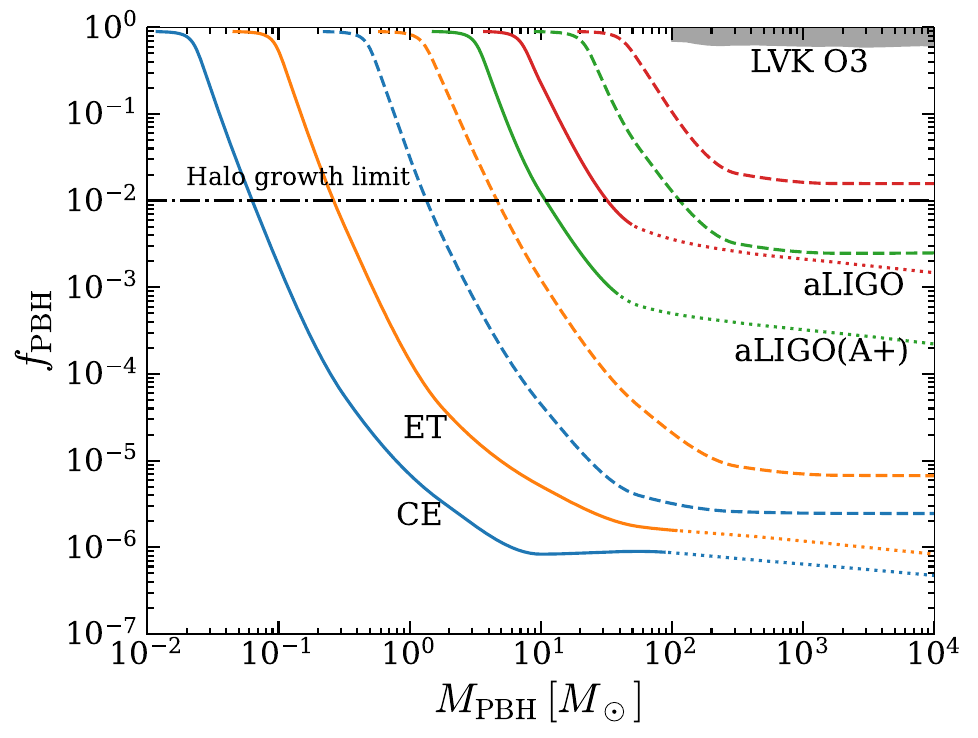} 
\caption{ \label{fig:result} 
Projected sensitivity on $f_{\rm PBH}$ at 90\% confidence level, assuming null GW lensing event detection for 5 yr of observations with aLIGO~\cite{abbott2020prospects}~(red curves), aLIGO+ (green curves), ET~\cite{hild2011sensitivity} (orange curves), and CE~\cite{srivastava2022science} (dark blue curves) experiments. dPBH results (solid curves) are compared with bare PBH results (dashed curves). Dotted curves that have the fraction of events allowing for discrimination of $\lesssim 10\%$ from \Fig{fig:prob}. Existing constraints from GW lensing searches of the LVK O3 phase are shown (gray shaded)~\cite{LIGOScientific:2021izm,LIGOScientific:2023bwz}. 
The results for $f_\text{PBH}\gtrsim 0.01$ (i.e., above the dot-dashed line) denoted by ``Halo growth limit'' can be affected by halo growth formation in a nontrivial way.}
\end{figure}

The expected 90\% upper limits on $f_{\rm PBH}$ are presented in \Fig{fig:result}, in the case of null detection over 5 years of observation. 
The occurrence of lensing events is assumed to follow a Poisson distribution with the above expectation value (see Supplemental Material). We consider the upcoming missions
aLIGO (design and A+ upgrade)~\cite{abbott2020prospects}, Cosmic Explorer (CE, low-frequency mode with 40 km arm length)~\cite{srivastava2022science} and Einstein Telescope (ET, ET-D configuration)~\cite{hild2011sensitivity} experiments.
Our reference results for bare PBHs agree with earlier studies~\cite{LIGOScientific:2023bwz,Basak:2021ten,Urrutia:2021qak}. Intriguingly, LVK O3 phase GW lensing observations started setting limits on PBHs contributing a sizable fraction of DM~\cite{LIGOScientific:2021izm,LIGOScientific:2023bwz}.
Other searches of bare PBHs (not shown) also constrain dPBHs of $\sim 10\textendash 10^2 M_{\odot}$ to be $f_{\rm PBH} \lesssim \mathcal{O}(10^{-3})$~\cite{Kohri:2018qtx,Ali-Haimoud:2016mbv,Zumalacarregui:2017qqd,Koushiappas:2017chw,Inoue:2017csr,Manshanden:2018tze,Lu:2020bmd,Serpico:2020ehh,Takhistov:2021aqx,Takhistov:2021upb,Manshanden:2018tze}. We do not discuss here in detail effects of dPBHs on different constraints. However, we note that unlike PBHs, dPBHs together with particle DM can contribute a significant fraction of DM abundance in this region.

In the near future, aLIGO will be able to probe stellar-mass PBHs below $f_{\rm PBH} \simeq 10^{-2}$, relevant for dPBHs. The sensitivity is about 2 orders of magnitudes better than that of current LVK O3 results with $\sim 70$ events. ET and CE can significantly further improve sensitivities, down to $f_{\rm PBH} \simeq 10^{-6}$ as well as extend into the subsolar mass range. The ET result is consistent with the recent dPBH study~\cite{Urrutia:2023mtk}.
Compared to bare PBH results, heavy-mass plateau regions, where dominant GO effects make $n_{\rm PBH} x_E^2$ not scale sensitively with $M_{\rm PBH}$, are improved by halo effects. The region also extends to lower masses as the dPBH scale is larger for the same $M_{\rm PBH}$.
Although GW lensing can become less sensitive to low-mass dPBHs as lensing shifts from the GO regime to the diffraction regime resulting in fewer detections, diffraction effects can increase the discrimination power per detection. 
When $f_\text{PBH}\gtrsim 0.01$ halo growth can be modified in a nontrivial way; hence, our results in this regime can be considered optimistic. 

\begin{figure}[t]
\includegraphics[width=0.99\linewidth,valign=t]{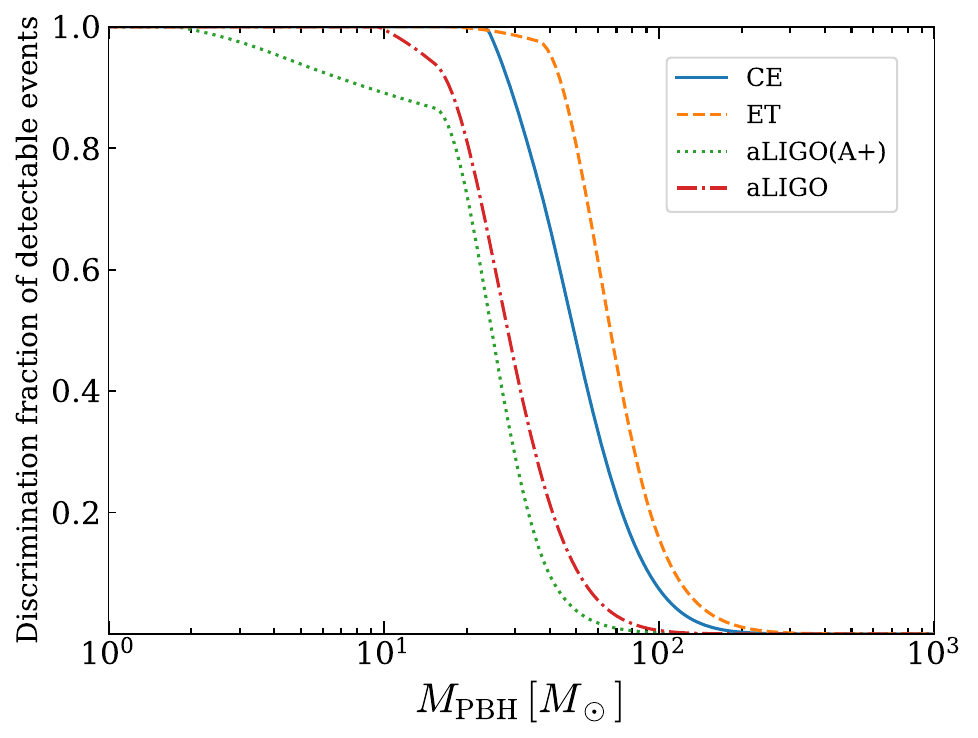} 
\caption{ \label{fig:prob} 
The fraction of total detectable lensed GW events that can allow discrimination of dPBH lenses against bare PBHs, with the SNR in the weak diffraction and bGO regimes $\geq 10$.}
\end{figure}

{\it Discrimination prospects}\textemdash  As discussed, discrimination is expected if both GO and diffraction regimes of the event are well measured. 
For effective discrimination, high-precision measurements of GW signals in the diffraction regime are crucial, while precision and the frequency range of the diffraction regime vary across events.
Thus, to estimate how often discrimination is possible, we calculate the fraction of detectable lensed events that have significant SNR $\geq 10$ in the weak diffraction and bGO regimes (see Supplemental Material). 
This fraction represents the portion of the total detectable GW lensing events allowing discrimination. We note that this quantity does not depend on $f_{\rm PBH}$.
As shown in \Fig{fig:prob}, this fraction is large for $M_{\rm PBH} \lesssim 10^2 \Msun$, signifying that the majority of lensed events are likely to have good discrimination prospects.
For heavier PBH lenses, GO effects are strong even for lower frequencies, resulting in small fraction of lensed events that allow discrimination.
Such regions with a low fraction $\leq 10\%$ of events allowing discrimination are also displayed as dotted curves in \Fig{fig:result}. 
Our calculations can be further improved in various ways, such as optimizing the SNR criteria in diffraction regimes. 
Our method is particularly promising for stellar-mass-range PBHs, relevant to LIGO's GW detection band.

{\it Conclusions}\textemdash 
LIGO-band GW measurements and theory of diffractive GW lensing are rapidly developing, allowing for unique novel insights into fundamental physics. We have shown that, when nontrivial amplification is observed over a large range of GW frequencies, matching analyses will be able to definitively distinguish dPBH from bare PBH lenses on an event-by-event basis. The events produced by lenses with mass $M_{\rm PBH} \lesssim 10^2 \Msun$ are particularly promising. Since such astrophysical BHs are not expected to form sizable DM halos, a single distinguishing event can imply the coexistence of stellar-mass PBHs and particle DM. These results open novel avenues for revealing the mysterious nature of DM, cosmological history, and the origin of merger GW events.

~\newline

{\it Acknowledgements}\textemdash 
We thank Kazunori Kohri, Masamune Oguri and Misao Sasaki for discussions. H.G.C. is supported by the Institute for Basic Science (IBS) under the project code, IBS-R018-D3. S.J and P.L. are supported by Grant Korea NRF2019R1C1C1010050.
V.T. acknowledges support by the World Premier International Research Center Initiative (WPI), MEXT, Japan and JSPS KAKENHI grant No. 23K13109.


\bibliography{references}

\begin{thebibliography}{108}%
\makeatletter
\providecommand \@ifxundefined [1]{%
 \@ifx{#1\undefined}
}%
\providecommand \@ifnum [1]{%
 \ifnum #1\expandafter \@firstoftwo
 \else \expandafter \@secondoftwo
 \fi
}%
\providecommand \@ifx [1]{%
 \ifx #1\expandafter \@firstoftwo
 \else \expandafter \@secondoftwo
 \fi
}%
\providecommand \natexlab [1]{#1}%
\providecommand \enquote  [1]{``#1''}%
\providecommand \bibnamefont  [1]{#1}%
\providecommand \bibfnamefont [1]{#1}%
\providecommand \citenamefont [1]{#1}%
\providecommand \href@noop [0]{\@secondoftwo}%
\providecommand \href [0]{\begingroup \@sanitize@url \@href}%
\providecommand \@href[1]{\@@startlink{#1}\@@href}%
\providecommand \@@href[1]{\endgroup#1\@@endlink}%
\providecommand \@sanitize@url [0]{\catcode `\\12\catcode `\$12\catcode `\&12\catcode `\#12\catcode `\^12\catcode `\_12\catcode `\%12\relax}%
\providecommand \@@startlink[1]{}%
\providecommand \@@endlink[0]{}%
\providecommand \url  [0]{\begingroup\@sanitize@url \@url }%
\providecommand \@url [1]{\endgroup\@href {#1}{\urlprefix }}%
\providecommand \urlprefix  [0]{URL }%
\providecommand \Eprint [0]{\href }%
\providecommand \doibase [0]{https://doi.org/}%
\providecommand \selectlanguage [0]{\@gobble}%
\providecommand \bibinfo  [0]{\@secondoftwo}%
\providecommand \bibfield  [0]{\@secondoftwo}%
\providecommand \translation [1]{[#1]}%
\providecommand \BibitemOpen [0]{}%
\providecommand \bibitemStop [0]{}%
\providecommand \bibitemNoStop [0]{.\EOS\space}%
\providecommand \EOS [0]{\spacefactor3000\relax}%
\providecommand \BibitemShut  [1]{\csname bibitem#1\endcsname}%
\let\auto@bib@innerbib\@empty
\bibitem [{\citenamefont {{Zel'dovich}}\ and\ \citenamefont {{Novikov}}(1967)}]{Zeldovich:1967}%
  \BibitemOpen
  \bibfield  {author} {\bibinfo {author} {\bibfnamefont {Y.~B.}\ \bibnamefont {{Zel'dovich}}}\ and\ \bibinfo {author} {\bibfnamefont {I.~D.}\ \bibnamefont {{Novikov}}},\ }\bibfield  {title} {\bibinfo {title} {{The Hypothesis of Cores Retarded during Expansion and the Hot Cosmological Model}},\ }\href@noop {} {\bibfield  {journal} {\bibinfo  {journal} {Sov. Astron.}\ }\textbf {\bibinfo {volume} {10}},\ \bibinfo {pages} {602} (\bibinfo {year} {1967})}\BibitemShut {NoStop}%
\bibitem [{\citenamefont {Hawking}(1971)}]{Hawking:1971ei}%
  \BibitemOpen
  \bibfield  {author} {\bibinfo {author} {\bibfnamefont {S.}~\bibnamefont {Hawking}},\ }\bibfield  {title} {\bibinfo {title} {{Gravitationally collapsed objects of very low mass}},\ }\href@noop {} {\bibfield  {journal} {\bibinfo  {journal} {Mon. Not. Roy. Astron. Soc.}\ }\textbf {\bibinfo {volume} {152}},\ \bibinfo {pages} {75} (\bibinfo {year} {1971})}\BibitemShut {NoStop}%
\bibitem [{\citenamefont {Carr}\ and\ \citenamefont {Hawking}(1974)}]{Carr:1974nx}%
  \BibitemOpen
  \bibfield  {author} {\bibinfo {author} {\bibfnamefont {B.~J.}\ \bibnamefont {Carr}}\ and\ \bibinfo {author} {\bibfnamefont {S.~W.}\ \bibnamefont {Hawking}},\ }\bibfield  {title} {\bibinfo {title} {{Black holes in the early Universe}},\ }\href {https://doi.org/10.1093/mnras/168.2.399} {\bibfield  {journal} {\bibinfo  {journal} {Mon. Not. Roy. Astron. Soc.}\ }\textbf {\bibinfo {volume} {168}},\ \bibinfo {pages} {399} (\bibinfo {year} {1974})}\BibitemShut {NoStop}%
\bibitem [{\citenamefont {Meszaros}(1975)}]{Meszaros:1975ef}%
  \BibitemOpen
  \bibfield  {author} {\bibinfo {author} {\bibfnamefont {P.}~\bibnamefont {Meszaros}},\ }\bibfield  {title} {\bibinfo {title} {{Primeval black holes and galaxy formation}},\ }\href@noop {} {\bibfield  {journal} {\bibinfo  {journal} {Astron. Astrophys.}\ }\textbf {\bibinfo {volume} {38}},\ \bibinfo {pages} {5} (\bibinfo {year} {1975})}\BibitemShut {NoStop}%
\bibitem [{\citenamefont {Carr}(1975)}]{Carr:1975qj}%
  \BibitemOpen
  \bibfield  {author} {\bibinfo {author} {\bibfnamefont {B.~J.}\ \bibnamefont {Carr}},\ }\bibfield  {title} {\bibinfo {title} {{The Primordial black hole mass spectrum}},\ }\href {https://doi.org/10.1086/153853} {\bibfield  {journal} {\bibinfo  {journal} {Astrophys. J.}\ }\textbf {\bibinfo {volume} {201}},\ \bibinfo {pages} {1} (\bibinfo {year} {1975})}\BibitemShut {NoStop}%
\bibitem [{\citenamefont {Garcia-Bellido}\ \emph {et~al.}(1996)\citenamefont {Garcia-Bellido}, \citenamefont {Linde},\ and\ \citenamefont {Wands}}]{GarciaBellido:1996qt}%
  \BibitemOpen
  \bibfield  {author} {\bibinfo {author} {\bibfnamefont {J.}~\bibnamefont {Garcia-Bellido}}, \bibinfo {author} {\bibfnamefont {A.~D.}\ \bibnamefont {Linde}},\ and\ \bibinfo {author} {\bibfnamefont {D.}~\bibnamefont {Wands}},\ }\bibfield  {title} {\bibinfo {title} {{Density perturbations and black hole formation in hybrid inflation}},\ }\href {https://doi.org/10.1103/PhysRevD.54.6040} {\bibfield  {journal} {\bibinfo  {journal} {Phys. Rev.}\ }\textbf {\bibinfo {volume} {D54}},\ \bibinfo {pages} {6040} (\bibinfo {year} {1996})},\ \Eprint {https://arxiv.org/abs/astro-ph/9605094} {arXiv:astro-ph/9605094 [astro-ph]} \BibitemShut {NoStop}%
\bibitem [{\citenamefont {Kawasaki}\ \emph {et~al.}(1998)\citenamefont {Kawasaki}, \citenamefont {Sugiyama},\ and\ \citenamefont {Yanagida}}]{Kawasaki:1997ju}%
  \BibitemOpen
  \bibfield  {author} {\bibinfo {author} {\bibfnamefont {M.}~\bibnamefont {Kawasaki}}, \bibinfo {author} {\bibfnamefont {N.}~\bibnamefont {Sugiyama}},\ and\ \bibinfo {author} {\bibfnamefont {T.}~\bibnamefont {Yanagida}},\ }\bibfield  {title} {\bibinfo {title} {{Primordial black hole formation in a double inflation model in supergravity}},\ }\href {https://doi.org/10.1103/PhysRevD.57.6050} {\bibfield  {journal} {\bibinfo  {journal} {Phys. Rev. D}\ }\textbf {\bibinfo {volume} {57}},\ \bibinfo {pages} {6050} (\bibinfo {year} {1998})},\ \Eprint {https://arxiv.org/abs/hep-ph/9710259} {arXiv:hep-ph/9710259} \BibitemShut {NoStop}%
\bibitem [{\citenamefont {Kohri}\ \emph {et~al.}(2008)\citenamefont {Kohri}, \citenamefont {Lyth},\ and\ \citenamefont {Melchiorri}}]{Kohri:2007qn}%
  \BibitemOpen
  \bibfield  {author} {\bibinfo {author} {\bibfnamefont {K.}~\bibnamefont {Kohri}}, \bibinfo {author} {\bibfnamefont {D.~H.}\ \bibnamefont {Lyth}},\ and\ \bibinfo {author} {\bibfnamefont {A.}~\bibnamefont {Melchiorri}},\ }\bibfield  {title} {\bibinfo {title} {{Black hole formation and slow-roll inflation}},\ }\href {https://doi.org/10.1088/1475-7516/2008/04/038} {\bibfield  {journal} {\bibinfo  {journal} {JCAP}\ }\textbf {\bibinfo {volume} {04}},\ \bibinfo {pages} {038}},\ \Eprint {https://arxiv.org/abs/0711.5006} {arXiv:0711.5006 [hep-ph]} \BibitemShut {NoStop}%
\bibitem [{\citenamefont {Khlopov}(2010)}]{Khlopov:2008qy}%
  \BibitemOpen
  \bibfield  {author} {\bibinfo {author} {\bibfnamefont {M.~Y.}\ \bibnamefont {Khlopov}},\ }\bibfield  {title} {\bibinfo {title} {{Primordial Black Holes}},\ }\href {https://doi.org/10.1088/1674-4527/10/6/001} {\bibfield  {journal} {\bibinfo  {journal} {Res. Astron. Astrophys.}\ }\textbf {\bibinfo {volume} {10}},\ \bibinfo {pages} {495} (\bibinfo {year} {2010})},\ \Eprint {https://arxiv.org/abs/0801.0116} {arXiv:0801.0116 [astro-ph]} \BibitemShut {NoStop}%
\bibitem [{\citenamefont {Frampton}\ \emph {et~al.}(2010)\citenamefont {Frampton}, \citenamefont {Kawasaki}, \citenamefont {Takahashi},\ and\ \citenamefont {Yanagida}}]{Frampton:2010sw}%
  \BibitemOpen
  \bibfield  {author} {\bibinfo {author} {\bibfnamefont {P.~H.}\ \bibnamefont {Frampton}}, \bibinfo {author} {\bibfnamefont {M.}~\bibnamefont {Kawasaki}}, \bibinfo {author} {\bibfnamefont {F.}~\bibnamefont {Takahashi}},\ and\ \bibinfo {author} {\bibfnamefont {T.~T.}\ \bibnamefont {Yanagida}},\ }\bibfield  {title} {\bibinfo {title} {{Primordial Black Holes as All Dark Matter}},\ }\href {https://doi.org/10.1088/1475-7516/2010/04/023} {\bibfield  {journal} {\bibinfo  {journal} {JCAP}\ }\textbf {\bibinfo {volume} {1004}},\ \bibinfo {pages} {023}},\ \Eprint {https://arxiv.org/abs/1001.2308} {arXiv:1001.2308 [hep-ph]} \BibitemShut {NoStop}%
\bibitem [{\citenamefont {Bird}\ \emph {et~al.}(2016)\citenamefont {Bird}, \citenamefont {Cholis}, \citenamefont {Mu\~noz}, \citenamefont {Ali-Ha\"\i{}moud}, \citenamefont {Kamionkowski}, \citenamefont {Kovetz}, \citenamefont {Raccanelli},\ and\ \citenamefont {Riess}}]{Bird:2016dcv}%
  \BibitemOpen
  \bibfield  {author} {\bibinfo {author} {\bibfnamefont {S.}~\bibnamefont {Bird}}, \bibinfo {author} {\bibfnamefont {I.}~\bibnamefont {Cholis}}, \bibinfo {author} {\bibfnamefont {J.~B.}\ \bibnamefont {Mu\~noz}}, \bibinfo {author} {\bibfnamefont {Y.}~\bibnamefont {Ali-Ha\"\i{}moud}}, \bibinfo {author} {\bibfnamefont {M.}~\bibnamefont {Kamionkowski}}, \bibinfo {author} {\bibfnamefont {E.~D.}\ \bibnamefont {Kovetz}}, \bibinfo {author} {\bibfnamefont {A.}~\bibnamefont {Raccanelli}},\ and\ \bibinfo {author} {\bibfnamefont {A.~G.}\ \bibnamefont {Riess}},\ }\bibfield  {title} {\bibinfo {title} {{Did LIGO detect dark matter?}},\ }\href {https://doi.org/10.1103/PhysRevLett.116.201301} {\bibfield  {journal} {\bibinfo  {journal} {Phys. Rev. Lett.}\ }\textbf {\bibinfo {volume} {116}},\ \bibinfo {pages} {201301} (\bibinfo {year} {2016})},\ \Eprint {https://arxiv.org/abs/1603.00464} {arXiv:1603.00464 [astro-ph.CO]} \BibitemShut {NoStop}%
\bibitem [{\citenamefont {Kawasaki}\ \emph {et~al.}(2016)\citenamefont {Kawasaki}, \citenamefont {Kusenko}, \citenamefont {Tada},\ and\ \citenamefont {Yanagida}}]{Kawasaki:2016pql}%
  \BibitemOpen
  \bibfield  {author} {\bibinfo {author} {\bibfnamefont {M.}~\bibnamefont {Kawasaki}}, \bibinfo {author} {\bibfnamefont {A.}~\bibnamefont {Kusenko}}, \bibinfo {author} {\bibfnamefont {Y.}~\bibnamefont {Tada}},\ and\ \bibinfo {author} {\bibfnamefont {T.~T.}\ \bibnamefont {Yanagida}},\ }\bibfield  {title} {\bibinfo {title} {{Primordial black holes as dark matter in supergravity inflation models}},\ }\href {https://doi.org/10.1103/PhysRevD.94.083523} {\bibfield  {journal} {\bibinfo  {journal} {Phys. Rev. D}\ }\textbf {\bibinfo {volume} {94}},\ \bibinfo {pages} {083523} (\bibinfo {year} {2016})},\ \Eprint {https://arxiv.org/abs/1606.07631} {arXiv:1606.07631 [astro-ph.CO]} \BibitemShut {NoStop}%
\bibitem [{\citenamefont {Inomata}\ \emph {et~al.}(2017)\citenamefont {Inomata}, \citenamefont {Kawasaki}, \citenamefont {Mukaida}, \citenamefont {Tada},\ and\ \citenamefont {Yanagida}}]{Inomata:2016rbd}%
  \BibitemOpen
  \bibfield  {author} {\bibinfo {author} {\bibfnamefont {K.}~\bibnamefont {Inomata}}, \bibinfo {author} {\bibfnamefont {M.}~\bibnamefont {Kawasaki}}, \bibinfo {author} {\bibfnamefont {K.}~\bibnamefont {Mukaida}}, \bibinfo {author} {\bibfnamefont {Y.}~\bibnamefont {Tada}},\ and\ \bibinfo {author} {\bibfnamefont {T.~T.}\ \bibnamefont {Yanagida}},\ }\bibfield  {title} {\bibinfo {title} {{Inflationary primordial black holes for the LIGO gravitational wave events and pulsar timing array experiments}},\ }\href {https://doi.org/10.1103/PhysRevD.95.123510} {\bibfield  {journal} {\bibinfo  {journal} {Phys. Rev. D}\ }\textbf {\bibinfo {volume} {95}},\ \bibinfo {pages} {123510} (\bibinfo {year} {2017})},\ \Eprint {https://arxiv.org/abs/1611.06130} {arXiv:1611.06130 [astro-ph.CO]} \BibitemShut {NoStop}%
\bibitem [{\citenamefont {Pi}\ \emph {et~al.}(2018)\citenamefont {Pi}, \citenamefont {Zhang}, \citenamefont {Huang},\ and\ \citenamefont {Sasaki}}]{Pi:2017gih}%
  \BibitemOpen
  \bibfield  {author} {\bibinfo {author} {\bibfnamefont {S.}~\bibnamefont {Pi}}, \bibinfo {author} {\bibfnamefont {Y.-l.}\ \bibnamefont {Zhang}}, \bibinfo {author} {\bibfnamefont {Q.-G.}\ \bibnamefont {Huang}},\ and\ \bibinfo {author} {\bibfnamefont {M.}~\bibnamefont {Sasaki}},\ }\bibfield  {title} {\bibinfo {title} {{Scalaron from $R^2$-gravity as a heavy field}},\ }\href {https://doi.org/10.1088/1475-7516/2018/05/042} {\bibfield  {journal} {\bibinfo  {journal} {JCAP}\ }\textbf {\bibinfo {volume} {05}},\ \bibinfo {pages} {042}},\ \Eprint {https://arxiv.org/abs/1712.09896} {arXiv:1712.09896 [astro-ph.CO]} \BibitemShut {NoStop}%
\bibitem [{\citenamefont {Garcia-Bellido}\ \emph {et~al.}(2017)\citenamefont {Garcia-Bellido}, \citenamefont {Peloso},\ and\ \citenamefont {Unal}}]{Garcia-Bellido:2017aan}%
  \BibitemOpen
  \bibfield  {author} {\bibinfo {author} {\bibfnamefont {J.}~\bibnamefont {Garcia-Bellido}}, \bibinfo {author} {\bibfnamefont {M.}~\bibnamefont {Peloso}},\ and\ \bibinfo {author} {\bibfnamefont {C.}~\bibnamefont {Unal}},\ }\bibfield  {title} {\bibinfo {title} {{Gravitational Wave signatures of inflationary models from Primordial Black Hole Dark Matter}},\ }\href {https://doi.org/10.1088/1475-7516/2017/09/013} {\bibfield  {journal} {\bibinfo  {journal} {JCAP}\ }\textbf {\bibinfo {volume} {09}},\ \bibinfo {pages} {013}},\ \Eprint {https://arxiv.org/abs/1707.02441} {arXiv:1707.02441 [astro-ph.CO]} \BibitemShut {NoStop}%
\bibitem [{\citenamefont {Georg}\ and\ \citenamefont {Watson}(2017)}]{Georg:2017mqk}%
  \BibitemOpen
  \bibfield  {author} {\bibinfo {author} {\bibfnamefont {J.}~\bibnamefont {Georg}}\ and\ \bibinfo {author} {\bibfnamefont {S.}~\bibnamefont {Watson}},\ }\bibfield  {title} {\bibinfo {title} {{A Preferred Mass Range for Primordial Black Hole Formation and Black Holes as Dark Matter Revisited}},\ }\href {https://doi.org/10.1007/JHEP09(2017)138} {\bibfield  {journal} {\bibinfo  {journal} {JHEP}\ }\textbf {\bibinfo {volume} {09}},\ \bibinfo {pages} {138}},\ \Eprint {https://arxiv.org/abs/1703.04825} {arXiv:1703.04825 [astro-ph.CO]} \BibitemShut {NoStop}%
\bibitem [{\citenamefont {Kocsis}\ \emph {et~al.}(2018)\citenamefont {Kocsis}, \citenamefont {Suyama}, \citenamefont {Tanaka},\ and\ \citenamefont {Yokoyama}}]{Kocsis:2017yty}%
  \BibitemOpen
  \bibfield  {author} {\bibinfo {author} {\bibfnamefont {B.}~\bibnamefont {Kocsis}}, \bibinfo {author} {\bibfnamefont {T.}~\bibnamefont {Suyama}}, \bibinfo {author} {\bibfnamefont {T.}~\bibnamefont {Tanaka}},\ and\ \bibinfo {author} {\bibfnamefont {S.}~\bibnamefont {Yokoyama}},\ }\bibfield  {title} {\bibinfo {title} {{Hidden universality in the merger rate distribution in the primordial black hole scenario}},\ }\href {https://doi.org/10.3847/1538-4357/aaa7f4} {\bibfield  {journal} {\bibinfo  {journal} {Astrophys. J.}\ }\textbf {\bibinfo {volume} {854}},\ \bibinfo {pages} {41} (\bibinfo {year} {2018})},\ \Eprint {https://arxiv.org/abs/1709.09007} {arXiv:1709.09007 [astro-ph.CO]} \BibitemShut {NoStop}%
\bibitem [{\citenamefont {Ando}\ \emph {et~al.}(2018)\citenamefont {Ando}, \citenamefont {Inomata}, \citenamefont {Kawasaki}, \citenamefont {Mukaida},\ and\ \citenamefont {Yanagida}}]{Ando:2017veq}%
  \BibitemOpen
  \bibfield  {author} {\bibinfo {author} {\bibfnamefont {K.}~\bibnamefont {Ando}}, \bibinfo {author} {\bibfnamefont {K.}~\bibnamefont {Inomata}}, \bibinfo {author} {\bibfnamefont {M.}~\bibnamefont {Kawasaki}}, \bibinfo {author} {\bibfnamefont {K.}~\bibnamefont {Mukaida}},\ and\ \bibinfo {author} {\bibfnamefont {T.~T.}\ \bibnamefont {Yanagida}},\ }\bibfield  {title} {\bibinfo {title} {{Primordial black holes for the LIGO events in the axionlike curvaton model}},\ }\href {https://doi.org/10.1103/PhysRevD.97.123512} {\bibfield  {journal} {\bibinfo  {journal} {Phys. Rev. D}\ }\textbf {\bibinfo {volume} {97}},\ \bibinfo {pages} {123512} (\bibinfo {year} {2018})},\ \Eprint {https://arxiv.org/abs/1711.08956} {arXiv:1711.08956 [astro-ph.CO]} \BibitemShut {NoStop}%
\bibitem [{\citenamefont {Cotner}\ and\ \citenamefont {Kusenko}(2017)}]{Cotner:2016cvr}%
  \BibitemOpen
  \bibfield  {author} {\bibinfo {author} {\bibfnamefont {E.}~\bibnamefont {Cotner}}\ and\ \bibinfo {author} {\bibfnamefont {A.}~\bibnamefont {Kusenko}},\ }\bibfield  {title} {\bibinfo {title} {{Primordial black holes from supersymmetry in the early universe}},\ }\href {https://doi.org/10.1103/PhysRevLett.119.031103} {\bibfield  {journal} {\bibinfo  {journal} {Phys. Rev. Lett.}\ }\textbf {\bibinfo {volume} {119}},\ \bibinfo {pages} {031103} (\bibinfo {year} {2017})},\ \Eprint {https://arxiv.org/abs/1612.02529} {arXiv:1612.02529 [astro-ph.CO]} \BibitemShut {NoStop}%
\bibitem [{\citenamefont {Cotner}\ \emph {et~al.}(2019)\citenamefont {Cotner}, \citenamefont {Kusenko}, \citenamefont {Sasaki},\ and\ \citenamefont {Takhistov}}]{Cotner:2019ykd}%
  \BibitemOpen
  \bibfield  {author} {\bibinfo {author} {\bibfnamefont {E.}~\bibnamefont {Cotner}}, \bibinfo {author} {\bibfnamefont {A.}~\bibnamefont {Kusenko}}, \bibinfo {author} {\bibfnamefont {M.}~\bibnamefont {Sasaki}},\ and\ \bibinfo {author} {\bibfnamefont {V.}~\bibnamefont {Takhistov}},\ }\bibfield  {title} {\bibinfo {title} {{Analytic Description of Primordial Black Hole Formation from Scalar Field Fragmentation}},\ }\href {https://doi.org/10.1088/1475-7516/2019/10/077} {\bibfield  {journal} {\bibinfo  {journal} {JCAP}\ }\textbf {\bibinfo {volume} {10}},\ \bibinfo {pages} {077}},\ \Eprint {https://arxiv.org/abs/1907.10613} {arXiv:1907.10613 [astro-ph.CO]} \BibitemShut {NoStop}%
\bibitem [{\citenamefont {Cotner}\ \emph {et~al.}(2018)\citenamefont {Cotner}, \citenamefont {Kusenko},\ and\ \citenamefont {Takhistov}}]{Cotner:2018vug}%
  \BibitemOpen
  \bibfield  {author} {\bibinfo {author} {\bibfnamefont {E.}~\bibnamefont {Cotner}}, \bibinfo {author} {\bibfnamefont {A.}~\bibnamefont {Kusenko}},\ and\ \bibinfo {author} {\bibfnamefont {V.}~\bibnamefont {Takhistov}},\ }\bibfield  {title} {\bibinfo {title} {{Primordial Black Holes from Inflaton Fragmentation into Oscillons}},\ }\href {https://doi.org/10.1103/PhysRevD.98.083513} {\bibfield  {journal} {\bibinfo  {journal} {Phys. Rev. D}\ }\textbf {\bibinfo {volume} {98}},\ \bibinfo {pages} {083513} (\bibinfo {year} {2018})},\ \Eprint {https://arxiv.org/abs/1801.03321} {arXiv:1801.03321 [astro-ph.CO]} \BibitemShut {NoStop}%
\bibitem [{\citenamefont {Sasaki}\ \emph {et~al.}(2018)\citenamefont {Sasaki}, \citenamefont {Suyama}, \citenamefont {Tanaka},\ and\ \citenamefont {Yokoyama}}]{Sasaki:2018dmp}%
  \BibitemOpen
  \bibfield  {author} {\bibinfo {author} {\bibfnamefont {M.}~\bibnamefont {Sasaki}}, \bibinfo {author} {\bibfnamefont {T.}~\bibnamefont {Suyama}}, \bibinfo {author} {\bibfnamefont {T.}~\bibnamefont {Tanaka}},\ and\ \bibinfo {author} {\bibfnamefont {S.}~\bibnamefont {Yokoyama}},\ }\bibfield  {title} {\bibinfo {title} {{Primordial black holes\textemdash{}perspectives in gravitational wave astronomy}},\ }\href {https://doi.org/10.1088/1361-6382/aaa7b4} {\bibfield  {journal} {\bibinfo  {journal} {Class. Quant. Grav.}\ }\textbf {\bibinfo {volume} {35}},\ \bibinfo {pages} {063001} (\bibinfo {year} {2018})},\ \Eprint {https://arxiv.org/abs/1801.05235} {arXiv:1801.05235 [astro-ph.CO]} \BibitemShut {NoStop}%
\bibitem [{\citenamefont {Carr}\ and\ \citenamefont {Silk}(2018)}]{Carr:2018rid}%
  \BibitemOpen
  \bibfield  {author} {\bibinfo {author} {\bibfnamefont {B.}~\bibnamefont {Carr}}\ and\ \bibinfo {author} {\bibfnamefont {J.}~\bibnamefont {Silk}},\ }\bibfield  {title} {\bibinfo {title} {{Primordial Black Holes as Generators of Cosmic Structures}},\ }\href {https://doi.org/10.1093/mnras/sty1204} {\bibfield  {journal} {\bibinfo  {journal} {Mon. Not. Roy. Astron. Soc.}\ }\textbf {\bibinfo {volume} {478}},\ \bibinfo {pages} {3756} (\bibinfo {year} {2018})},\ \Eprint {https://arxiv.org/abs/1801.00672} {arXiv:1801.00672 [astro-ph.CO]} \BibitemShut {NoStop}%
\bibitem [{\citenamefont {Kusenko}\ \emph {et~al.}(2020)\citenamefont {Kusenko}, \citenamefont {Sasaki}, \citenamefont {Sugiyama}, \citenamefont {Takada}, \citenamefont {Takhistov},\ and\ \citenamefont {Vitagliano}}]{Kusenko:2020pcg}%
  \BibitemOpen
  \bibfield  {author} {\bibinfo {author} {\bibfnamefont {A.}~\bibnamefont {Kusenko}}, \bibinfo {author} {\bibfnamefont {M.}~\bibnamefont {Sasaki}}, \bibinfo {author} {\bibfnamefont {S.}~\bibnamefont {Sugiyama}}, \bibinfo {author} {\bibfnamefont {M.}~\bibnamefont {Takada}}, \bibinfo {author} {\bibfnamefont {V.}~\bibnamefont {Takhistov}},\ and\ \bibinfo {author} {\bibfnamefont {E.}~\bibnamefont {Vitagliano}},\ }\bibfield  {title} {\bibinfo {title} {{Exploring Primordial Black Holes from the Multiverse with Optical Telescopes}},\ }\href {https://doi.org/10.1103/PhysRevLett.125.181304} {\bibfield  {journal} {\bibinfo  {journal} {Phys. Rev. Lett.}\ }\textbf {\bibinfo {volume} {125}},\ \bibinfo {pages} {181304} (\bibinfo {year} {2020})},\ \Eprint {https://arxiv.org/abs/2001.09160} {arXiv:2001.09160 [astro-ph.CO]} \BibitemShut {NoStop}%
\bibitem [{\citenamefont {Carr}\ \emph {et~al.}(2021)\citenamefont {Carr}, \citenamefont {Kohri}, \citenamefont {Sendouda},\ and\ \citenamefont {Yokoyama}}]{Carr:2020gox}%
  \BibitemOpen
  \bibfield  {author} {\bibinfo {author} {\bibfnamefont {B.}~\bibnamefont {Carr}}, \bibinfo {author} {\bibfnamefont {K.}~\bibnamefont {Kohri}}, \bibinfo {author} {\bibfnamefont {Y.}~\bibnamefont {Sendouda}},\ and\ \bibinfo {author} {\bibfnamefont {J.}~\bibnamefont {Yokoyama}},\ }\bibfield  {title} {\bibinfo {title} {{Constraints on primordial black holes}},\ }\href {https://doi.org/10.1088/1361-6633/ac1e31} {\bibfield  {journal} {\bibinfo  {journal} {Rept. Prog. Phys.}\ }\textbf {\bibinfo {volume} {84}},\ \bibinfo {pages} {116902} (\bibinfo {year} {2021})},\ \Eprint {https://arxiv.org/abs/2002.12778} {arXiv:2002.12778 [astro-ph.CO]} \BibitemShut {NoStop}%
\bibitem [{\citenamefont {Green}\ and\ \citenamefont {Kavanagh}(2021)}]{Green:2020jor}%
  \BibitemOpen
  \bibfield  {author} {\bibinfo {author} {\bibfnamefont {A.~M.}\ \bibnamefont {Green}}\ and\ \bibinfo {author} {\bibfnamefont {B.~J.}\ \bibnamefont {Kavanagh}},\ }\bibfield  {title} {\bibinfo {title} {{Primordial Black Holes as a dark matter candidate}},\ }\href {https://doi.org/10.1088/1361-6471/abc534} {\bibfield  {journal} {\bibinfo  {journal} {J. Phys. G}\ }\textbf {\bibinfo {volume} {48}},\ \bibinfo {pages} {043001} (\bibinfo {year} {2021})},\ \Eprint {https://arxiv.org/abs/2007.10722} {arXiv:2007.10722 [astro-ph.CO]} \BibitemShut {NoStop}%
\bibitem [{\citenamefont {Escriva}\ \emph {et~al.}(2022)\citenamefont {Escriva}, \citenamefont {Kuhnel},\ and\ \citenamefont {Tada}}]{Escriva:2022duf}%
  \BibitemOpen
  \bibfield  {author} {\bibinfo {author} {\bibfnamefont {A.}~\bibnamefont {Escriva}}, \bibinfo {author} {\bibfnamefont {F.}~\bibnamefont {Kuhnel}},\ and\ \bibinfo {author} {\bibfnamefont {Y.}~\bibnamefont {Tada}},\ }\bibfield  {title} {\bibinfo {title} {{Primordial Black Holes}},\ }\href@noop {} {\  (\bibinfo {year} {2022})},\ \Eprint {https://arxiv.org/abs/2211.05767} {arXiv:2211.05767 [astro-ph.CO]} \BibitemShut {NoStop}%
\bibitem [{\citenamefont {Kawana}\ and\ \citenamefont {Xie}(2022)}]{Kawana:2021tde}%
  \BibitemOpen
  \bibfield  {author} {\bibinfo {author} {\bibfnamefont {K.}~\bibnamefont {Kawana}}\ and\ \bibinfo {author} {\bibfnamefont {K.-P.}\ \bibnamefont {Xie}},\ }\bibfield  {title} {\bibinfo {title} {{Primordial black holes from a cosmic phase transition: The collapse of Fermi-balls}},\ }\href {https://doi.org/10.1016/j.physletb.2021.136791} {\bibfield  {journal} {\bibinfo  {journal} {Phys. Lett. B}\ }\textbf {\bibinfo {volume} {824}},\ \bibinfo {pages} {136791} (\bibinfo {year} {2022})},\ \Eprint {https://arxiv.org/abs/2106.00111} {arXiv:2106.00111 [astro-ph.CO]} \BibitemShut {NoStop}%
\bibitem [{\citenamefont {Lu}\ \emph {et~al.}(2023{\natexlab{a}})\citenamefont {Lu}, \citenamefont {Kawana},\ and\ \citenamefont {Kusenko}}]{Lu:2022jnp}%
  \BibitemOpen
  \bibfield  {author} {\bibinfo {author} {\bibfnamefont {P.}~\bibnamefont {Lu}}, \bibinfo {author} {\bibfnamefont {K.}~\bibnamefont {Kawana}},\ and\ \bibinfo {author} {\bibfnamefont {A.}~\bibnamefont {Kusenko}},\ }\bibfield  {title} {\bibinfo {title} {{Late-forming primordial black holes: Beyond the CMB era}},\ }\href {https://doi.org/10.1103/PhysRevD.107.103037} {\bibfield  {journal} {\bibinfo  {journal} {Phys. Rev. D}\ }\textbf {\bibinfo {volume} {107}},\ \bibinfo {pages} {103037} (\bibinfo {year} {2023}{\natexlab{a}})},\ \Eprint {https://arxiv.org/abs/2210.16462} {arXiv:2210.16462 [astro-ph.CO]} \BibitemShut {NoStop}%
\bibitem [{\citenamefont {Kawana}\ \emph {et~al.}(2022)\citenamefont {Kawana}, \citenamefont {Lu},\ and\ \citenamefont {Xie}}]{Kawana:2022lba}%
  \BibitemOpen
  \bibfield  {author} {\bibinfo {author} {\bibfnamefont {K.}~\bibnamefont {Kawana}}, \bibinfo {author} {\bibfnamefont {P.}~\bibnamefont {Lu}},\ and\ \bibinfo {author} {\bibfnamefont {K.-P.}\ \bibnamefont {Xie}},\ }\bibfield  {title} {\bibinfo {title} {{First-order phase transition and fate of false vacuum remnants}},\ }\href {https://doi.org/10.1088/1475-7516/2022/10/030} {\bibfield  {journal} {\bibinfo  {journal} {JCAP}\ }\textbf {\bibinfo {volume} {10}},\ \bibinfo {pages} {030}},\ \Eprint {https://arxiv.org/abs/2206.09923} {arXiv:2206.09923 [astro-ph.CO]} \BibitemShut {NoStop}%
\bibitem [{\citenamefont {Lu}\ \emph {et~al.}(2022)\citenamefont {Lu}, \citenamefont {Kawana},\ and\ \citenamefont {Xie}}]{Lu:2022paj}%
  \BibitemOpen
  \bibfield  {author} {\bibinfo {author} {\bibfnamefont {P.}~\bibnamefont {Lu}}, \bibinfo {author} {\bibfnamefont {K.}~\bibnamefont {Kawana}},\ and\ \bibinfo {author} {\bibfnamefont {K.-P.}\ \bibnamefont {Xie}},\ }\bibfield  {title} {\bibinfo {title} {{Old phase remnants in first-order phase transitions}},\ }\href {https://doi.org/10.1103/PhysRevD.105.123503} {\bibfield  {journal} {\bibinfo  {journal} {Phys. Rev. D}\ }\textbf {\bibinfo {volume} {105}},\ \bibinfo {pages} {123503} (\bibinfo {year} {2022})},\ \Eprint {https://arxiv.org/abs/2202.03439} {arXiv:2202.03439 [astro-ph.CO]} \BibitemShut {NoStop}%
\bibitem [{\citenamefont {Lu}\ \emph {et~al.}(2023{\natexlab{b}})\citenamefont {Lu}, \citenamefont {Takhistov},\ and\ \citenamefont {Fuller}}]{Lu:2022yuc}%
  \BibitemOpen
  \bibfield  {author} {\bibinfo {author} {\bibfnamefont {P.}~\bibnamefont {Lu}}, \bibinfo {author} {\bibfnamefont {V.}~\bibnamefont {Takhistov}},\ and\ \bibinfo {author} {\bibfnamefont {G.~M.}\ \bibnamefont {Fuller}},\ }\bibfield  {title} {\bibinfo {title} {{Signatures of a High Temperature QCD Transition in the Early Universe}},\ }\href {https://doi.org/10.1103/PhysRevLett.130.221002} {\bibfield  {journal} {\bibinfo  {journal} {Phys. Rev. Lett.}\ }\textbf {\bibinfo {volume} {130}},\ \bibinfo {pages} {221002} (\bibinfo {year} {2023}{\natexlab{b}})},\ \Eprint {https://arxiv.org/abs/2212.00156} {arXiv:2212.00156 [astro-ph.CO]} \BibitemShut {NoStop}%
\bibitem [{\citenamefont {Kawana}\ \emph {et~al.}(2023)\citenamefont {Kawana}, \citenamefont {Kim},\ and\ \citenamefont {Lu}}]{Kawana:2022olo}%
  \BibitemOpen
  \bibfield  {author} {\bibinfo {author} {\bibfnamefont {K.}~\bibnamefont {Kawana}}, \bibinfo {author} {\bibfnamefont {T.}~\bibnamefont {Kim}},\ and\ \bibinfo {author} {\bibfnamefont {P.}~\bibnamefont {Lu}},\ }\bibfield  {title} {\bibinfo {title} {{PBH formation from overdensities in delayed vacuum transitions}},\ }\href {https://doi.org/10.1103/PhysRevD.108.103531} {\bibfield  {journal} {\bibinfo  {journal} {Phys. Rev. D}\ }\textbf {\bibinfo {volume} {108}},\ \bibinfo {pages} {103531} (\bibinfo {year} {2023})},\ \Eprint {https://arxiv.org/abs/2212.14037} {arXiv:2212.14037 [astro-ph.CO]} \BibitemShut {NoStop}%
\bibitem [{\citenamefont {Chakraborty}\ \emph {et~al.}(2022)\citenamefont {Chakraborty}, \citenamefont {Chanda}, \citenamefont {Pandey},\ and\ \citenamefont {Das}}]{Chakraborty:2022mwu}%
  \BibitemOpen
  \bibfield  {author} {\bibinfo {author} {\bibfnamefont {A.}~\bibnamefont {Chakraborty}}, \bibinfo {author} {\bibfnamefont {P.~K.}\ \bibnamefont {Chanda}}, \bibinfo {author} {\bibfnamefont {K.~L.}\ \bibnamefont {Pandey}},\ and\ \bibinfo {author} {\bibfnamefont {S.}~\bibnamefont {Das}},\ }\bibfield  {title} {\bibinfo {title} {{Formation and Abundance of Late-forming Primordial Black Holes as Dark Matter}},\ }\href {https://doi.org/10.3847/1538-4357/ac6ddd} {\bibfield  {journal} {\bibinfo  {journal} {Astrophys. J.}\ }\textbf {\bibinfo {volume} {932}},\ \bibinfo {pages} {119} (\bibinfo {year} {2022})},\ \Eprint {https://arxiv.org/abs/2204.09628} {arXiv:2204.09628 [astro-ph.CO]} \BibitemShut {NoStop}%
\bibitem [{\citenamefont {Kohri}\ and\ \citenamefont {Terada}(2018)}]{Kohri:2018qtx}%
  \BibitemOpen
  \bibfield  {author} {\bibinfo {author} {\bibfnamefont {K.}~\bibnamefont {Kohri}}\ and\ \bibinfo {author} {\bibfnamefont {T.}~\bibnamefont {Terada}},\ }\bibfield  {title} {\bibinfo {title} {{Primordial Black Hole Dark Matter and LIGO/Virgo Merger Rate from Inflation with Running Spectral Indices: Formation in the Matter- and/or Radiation-Dominated Universe}},\ }\href {https://doi.org/10.1088/1361-6382/aaea18} {\bibfield  {journal} {\bibinfo  {journal} {Class. Quant. Grav.}\ }\textbf {\bibinfo {volume} {35}},\ \bibinfo {pages} {235017} (\bibinfo {year} {2018})},\ \Eprint {https://arxiv.org/abs/1802.06785} {arXiv:1802.06785 [astro-ph.CO]} \BibitemShut {NoStop}%
\bibitem [{\citenamefont {Ali-Ha{\"\i}moud}\ and\ \citenamefont {Kamionkowski}(2017)}]{Ali-Haimoud:2016mbv}%
  \BibitemOpen
  \bibfield  {author} {\bibinfo {author} {\bibfnamefont {Y.}~\bibnamefont {Ali-Ha{\"\i}moud}}\ and\ \bibinfo {author} {\bibfnamefont {M.}~\bibnamefont {Kamionkowski}},\ }\bibfield  {title} {\bibinfo {title} {{Cosmic microwave background limits on accreting primordial black holes}},\ }\href {https://doi.org/10.1103/PhysRevD.95.043534} {\bibfield  {journal} {\bibinfo  {journal} {Phys. Rev.}\ }\textbf {\bibinfo {volume} {D95}},\ \bibinfo {pages} {043534} (\bibinfo {year} {2017})},\ \Eprint {https://arxiv.org/abs/1612.05644} {arXiv:1612.05644 [astro-ph.CO]} \BibitemShut {NoStop}%
\bibitem [{\citenamefont {Zumalacarregui}\ and\ \citenamefont {Seljak}(2018)}]{Zumalacarregui:2017qqd}%
  \BibitemOpen
  \bibfield  {author} {\bibinfo {author} {\bibfnamefont {M.}~\bibnamefont {Zumalacarregui}}\ and\ \bibinfo {author} {\bibfnamefont {U.}~\bibnamefont {Seljak}},\ }\bibfield  {title} {\bibinfo {title} {{Limits on stellar-mass compact objects as dark matter from gravitational lensing of type Ia supernovae}},\ }\href {https://doi.org/10.1103/PhysRevLett.121.141101} {\bibfield  {journal} {\bibinfo  {journal} {Phys. Rev. Lett.}\ }\textbf {\bibinfo {volume} {121}},\ \bibinfo {pages} {141101} (\bibinfo {year} {2018})},\ \Eprint {https://arxiv.org/abs/1712.02240} {arXiv:1712.02240 [astro-ph.CO]} \BibitemShut {NoStop}%
\bibitem [{\citenamefont {Koushiappas}\ and\ \citenamefont {Loeb}(2017)}]{Koushiappas:2017chw}%
  \BibitemOpen
  \bibfield  {author} {\bibinfo {author} {\bibfnamefont {S.~M.}\ \bibnamefont {Koushiappas}}\ and\ \bibinfo {author} {\bibfnamefont {A.}~\bibnamefont {Loeb}},\ }\bibfield  {title} {\bibinfo {title} {{Dynamics of Dwarf Galaxies Disfavor Stellar-Mass Black Holes as Dark Matter}},\ }\href {https://doi.org/10.1103/PhysRevLett.119.041102} {\bibfield  {journal} {\bibinfo  {journal} {Phys. Rev. Lett.}\ }\textbf {\bibinfo {volume} {119}},\ \bibinfo {pages} {041102} (\bibinfo {year} {2017})},\ \Eprint {https://arxiv.org/abs/1704.01668} {arXiv:1704.01668 [astro-ph.GA]} \BibitemShut {NoStop}%
\bibitem [{\citenamefont {Inoue}\ and\ \citenamefont {Kusenko}(2017)}]{Inoue:2017csr}%
  \BibitemOpen
  \bibfield  {author} {\bibinfo {author} {\bibfnamefont {Y.}~\bibnamefont {Inoue}}\ and\ \bibinfo {author} {\bibfnamefont {A.}~\bibnamefont {Kusenko}},\ }\bibfield  {title} {\bibinfo {title} {{New X-ray bound on density of primordial black holes}},\ }\href {https://doi.org/10.1088/1475-7516/2017/10/034} {\bibfield  {journal} {\bibinfo  {journal} {JCAP}\ }\textbf {\bibinfo {volume} {1710}}\bibfield  {number} {\bibinfo  {number} { (10)},\ \bibinfo {pages} {034}},\ }\Eprint {https://arxiv.org/abs/1705.00791} {arXiv:1705.00791 [astro-ph.CO]} \BibitemShut {NoStop}%
\bibitem [{\citenamefont {Manshanden}\ \emph {et~al.}(2019)\citenamefont {Manshanden}, \citenamefont {Gaggero}, \citenamefont {Bertone}, \citenamefont {Connors},\ and\ \citenamefont {Ricotti}}]{Manshanden:2018tze}%
  \BibitemOpen
  \bibfield  {author} {\bibinfo {author} {\bibfnamefont {J.}~\bibnamefont {Manshanden}}, \bibinfo {author} {\bibfnamefont {D.}~\bibnamefont {Gaggero}}, \bibinfo {author} {\bibfnamefont {G.}~\bibnamefont {Bertone}}, \bibinfo {author} {\bibfnamefont {R.~M.~T.}\ \bibnamefont {Connors}},\ and\ \bibinfo {author} {\bibfnamefont {M.}~\bibnamefont {Ricotti}},\ }\bibfield  {title} {\bibinfo {title} {{Multi-wavelength astronomical searches for primordial black holes}},\ }\href {https://doi.org/10.1088/1475-7516/2019/06/026} {\bibfield  {journal} {\bibinfo  {journal} {JCAP}\ }\textbf {\bibinfo {volume} {06}},\ \bibinfo {pages} {026}},\ \Eprint {https://arxiv.org/abs/1812.07967} {arXiv:1812.07967 [astro-ph.HE]} \BibitemShut {NoStop}%
\bibitem [{\citenamefont {Lu}\ \emph {et~al.}(2021)\citenamefont {Lu}, \citenamefont {Takhistov}, \citenamefont {Gelmini}, \citenamefont {Hayashi}, \citenamefont {Inoue},\ and\ \citenamefont {Kusenko}}]{Lu:2020bmd}%
  \BibitemOpen
  \bibfield  {author} {\bibinfo {author} {\bibfnamefont {P.}~\bibnamefont {Lu}}, \bibinfo {author} {\bibfnamefont {V.}~\bibnamefont {Takhistov}}, \bibinfo {author} {\bibfnamefont {G.~B.}\ \bibnamefont {Gelmini}}, \bibinfo {author} {\bibfnamefont {K.}~\bibnamefont {Hayashi}}, \bibinfo {author} {\bibfnamefont {Y.}~\bibnamefont {Inoue}},\ and\ \bibinfo {author} {\bibfnamefont {A.}~\bibnamefont {Kusenko}},\ }\bibfield  {title} {\bibinfo {title} {{Constraining Primordial Black Holes with Dwarf Galaxy Heating}},\ }\href {https://doi.org/10.3847/2041-8213/abdcb6} {\bibfield  {journal} {\bibinfo  {journal} {Astrophys. J. Lett.}\ }\textbf {\bibinfo {volume} {908}},\ \bibinfo {pages} {L23} (\bibinfo {year} {2021})},\ \Eprint {https://arxiv.org/abs/2007.02213} {arXiv:2007.02213 [astro-ph.CO]} \BibitemShut {NoStop}%
\bibitem [{\citenamefont {Serpico}\ \emph {et~al.}(2020)\citenamefont {Serpico}, \citenamefont {Poulin}, \citenamefont {Inman},\ and\ \citenamefont {Kohri}}]{Serpico:2020ehh}%
  \BibitemOpen
  \bibfield  {author} {\bibinfo {author} {\bibfnamefont {P.~D.}\ \bibnamefont {Serpico}}, \bibinfo {author} {\bibfnamefont {V.}~\bibnamefont {Poulin}}, \bibinfo {author} {\bibfnamefont {D.}~\bibnamefont {Inman}},\ and\ \bibinfo {author} {\bibfnamefont {K.}~\bibnamefont {Kohri}},\ }\bibfield  {title} {\bibinfo {title} {{Cosmic microwave background bounds on primordial black holes including dark matter halo accretion}},\ }\href {https://doi.org/10.1103/PhysRevResearch.2.023204} {\bibfield  {journal} {\bibinfo  {journal} {Phys. Rev. Res.}\ }\textbf {\bibinfo {volume} {2}},\ \bibinfo {pages} {023204} (\bibinfo {year} {2020})},\ \Eprint {https://arxiv.org/abs/2002.10771} {arXiv:2002.10771 [astro-ph.CO]} \BibitemShut {NoStop}%
\bibitem [{\citenamefont {Takhistov}\ \emph {et~al.}(2022{\natexlab{a}})\citenamefont {Takhistov}, \citenamefont {Lu}, \citenamefont {Gelmini}, \citenamefont {Hayashi}, \citenamefont {Inoue},\ and\ \citenamefont {Kusenko}}]{Takhistov:2021aqx}%
  \BibitemOpen
  \bibfield  {author} {\bibinfo {author} {\bibfnamefont {V.}~\bibnamefont {Takhistov}}, \bibinfo {author} {\bibfnamefont {P.}~\bibnamefont {Lu}}, \bibinfo {author} {\bibfnamefont {G.~B.}\ \bibnamefont {Gelmini}}, \bibinfo {author} {\bibfnamefont {K.}~\bibnamefont {Hayashi}}, \bibinfo {author} {\bibfnamefont {Y.}~\bibnamefont {Inoue}},\ and\ \bibinfo {author} {\bibfnamefont {A.}~\bibnamefont {Kusenko}},\ }\bibfield  {title} {\bibinfo {title} {{Interstellar gas heating by primordial black holes}},\ }\href {https://doi.org/10.1088/1475-7516/2022/03/017} {\bibfield  {journal} {\bibinfo  {journal} {JCAP}\ }\textbf {\bibinfo {volume} {03}}\bibfield  {number} {\bibinfo  {number} { (03)},\ \bibinfo {pages} {017}},\ }\Eprint {https://arxiv.org/abs/2105.06099} {arXiv:2105.06099 [astro-ph.GA]} \BibitemShut {NoStop}%
\bibitem [{\citenamefont {Takhistov}\ \emph {et~al.}(2022{\natexlab{b}})\citenamefont {Takhistov}, \citenamefont {Lu}, \citenamefont {Murase}, \citenamefont {Inoue},\ and\ \citenamefont {Gelmini}}]{Takhistov:2021upb}%
  \BibitemOpen
  \bibfield  {author} {\bibinfo {author} {\bibfnamefont {V.}~\bibnamefont {Takhistov}}, \bibinfo {author} {\bibfnamefont {P.}~\bibnamefont {Lu}}, \bibinfo {author} {\bibfnamefont {K.}~\bibnamefont {Murase}}, \bibinfo {author} {\bibfnamefont {Y.}~\bibnamefont {Inoue}},\ and\ \bibinfo {author} {\bibfnamefont {G.~B.}\ \bibnamefont {Gelmini}},\ }\bibfield  {title} {\bibinfo {title} {{Impacts of Jets and winds from primordial black holes}},\ }\href {https://doi.org/10.1093/mnrasl/slac097} {\bibfield  {journal} {\bibinfo  {journal} {Mon. Not. Roy. Astron. Soc.}\ }\textbf {\bibinfo {volume} {517}},\ \bibinfo {pages} {L1} (\bibinfo {year} {2022}{\natexlab{b}})},\ \Eprint {https://arxiv.org/abs/2111.08699} {arXiv:2111.08699 [astro-ph.HE]} \BibitemShut {NoStop}%
\bibitem [{\citenamefont {Clesse}\ and\ \citenamefont {Garc\'\i{}a-Bellido}(2017)}]{Clesse:2016vqa}%
  \BibitemOpen
  \bibfield  {author} {\bibinfo {author} {\bibfnamefont {S.}~\bibnamefont {Clesse}}\ and\ \bibinfo {author} {\bibfnamefont {J.}~\bibnamefont {Garc\'\i{}a-Bellido}},\ }\bibfield  {title} {\bibinfo {title} {{The clustering of massive Primordial Black Holes as Dark Matter: measuring their mass distribution with Advanced LIGO}},\ }\href {https://doi.org/10.1016/j.dark.2016.10.002} {\bibfield  {journal} {\bibinfo  {journal} {Phys. Dark Univ.}\ }\textbf {\bibinfo {volume} {15}},\ \bibinfo {pages} {142} (\bibinfo {year} {2017})},\ \Eprint {https://arxiv.org/abs/1603.05234} {arXiv:1603.05234 [astro-ph.CO]} \BibitemShut {NoStop}%
\bibitem [{\citenamefont {Sasaki}\ \emph {et~al.}(2016)\citenamefont {Sasaki}, \citenamefont {Suyama}, \citenamefont {Tanaka},\ and\ \citenamefont {Yokoyama}}]{Sasaki:2016jop}%
  \BibitemOpen
  \bibfield  {author} {\bibinfo {author} {\bibfnamefont {M.}~\bibnamefont {Sasaki}}, \bibinfo {author} {\bibfnamefont {T.}~\bibnamefont {Suyama}}, \bibinfo {author} {\bibfnamefont {T.}~\bibnamefont {Tanaka}},\ and\ \bibinfo {author} {\bibfnamefont {S.}~\bibnamefont {Yokoyama}},\ }\bibfield  {title} {\bibinfo {title} {{Primordial Black Hole Scenario for the Gravitational-Wave Event GW150914}},\ }\href {https://doi.org/10.1103/PhysRevLett.117.061101} {\bibfield  {journal} {\bibinfo  {journal} {Phys. Rev. Lett.}\ }\textbf {\bibinfo {volume} {117}},\ \bibinfo {pages} {061101} (\bibinfo {year} {2016})},\ \bibinfo {note} {[Erratum: Phys.Rev.Lett. 121, 059901 (2018)]},\ \Eprint {https://arxiv.org/abs/1603.08338} {arXiv:1603.08338 [astro-ph.CO]} \BibitemShut {NoStop}%
\bibitem [{\citenamefont {Vaskonen}\ and\ \citenamefont {Veerm\"ae}(2020)}]{Vaskonen:2019jpv}%
  \BibitemOpen
  \bibfield  {author} {\bibinfo {author} {\bibfnamefont {V.}~\bibnamefont {Vaskonen}}\ and\ \bibinfo {author} {\bibfnamefont {H.}~\bibnamefont {Veerm\"ae}},\ }\bibfield  {title} {\bibinfo {title} {{Lower bound on the primordial black hole merger rate}},\ }\href {https://doi.org/10.1103/PhysRevD.101.043015} {\bibfield  {journal} {\bibinfo  {journal} {Phys. Rev. D}\ }\textbf {\bibinfo {volume} {101}},\ \bibinfo {pages} {043015} (\bibinfo {year} {2020})},\ \Eprint {https://arxiv.org/abs/1908.09752} {arXiv:1908.09752 [astro-ph.CO]} \BibitemShut {NoStop}%
\bibitem [{\citenamefont {Franciolini}\ \emph {et~al.}(2022)\citenamefont {Franciolini}, \citenamefont {Baibhav}, \citenamefont {De~Luca}, \citenamefont {Ng}, \citenamefont {Wong}, \citenamefont {Berti}, \citenamefont {Pani}, \citenamefont {Riotto},\ and\ \citenamefont {Vitale}}]{Franciolini:2021tla}%
  \BibitemOpen
  \bibfield  {author} {\bibinfo {author} {\bibfnamefont {G.}~\bibnamefont {Franciolini}}, \bibinfo {author} {\bibfnamefont {V.}~\bibnamefont {Baibhav}}, \bibinfo {author} {\bibfnamefont {V.}~\bibnamefont {De~Luca}}, \bibinfo {author} {\bibfnamefont {K.~K.~Y.}\ \bibnamefont {Ng}}, \bibinfo {author} {\bibfnamefont {K.~W.~K.}\ \bibnamefont {Wong}}, \bibinfo {author} {\bibfnamefont {E.}~\bibnamefont {Berti}}, \bibinfo {author} {\bibfnamefont {P.}~\bibnamefont {Pani}}, \bibinfo {author} {\bibfnamefont {A.}~\bibnamefont {Riotto}},\ and\ \bibinfo {author} {\bibfnamefont {S.}~\bibnamefont {Vitale}},\ }\bibfield  {title} {\bibinfo {title} {{Searching for a subpopulation of primordial black holes in LIGO-Virgo gravitational-wave data}},\ }\href {https://doi.org/10.1103/PhysRevD.105.083526} {\bibfield  {journal} {\bibinfo  {journal} {Phys. Rev. D}\ }\textbf {\bibinfo {volume} {105}},\ \bibinfo {pages} {083526} (\bibinfo {year} {2022})},\ \Eprint {https://arxiv.org/abs/2105.03349} {arXiv:2105.03349 [gr-qc]} \BibitemShut
  {NoStop}%
\bibitem [{\citenamefont {Fuller}\ \emph {et~al.}(2017)\citenamefont {Fuller}, \citenamefont {Kusenko},\ and\ \citenamefont {Takhistov}}]{Fuller:2017uyd}%
  \BibitemOpen
  \bibfield  {author} {\bibinfo {author} {\bibfnamefont {G.~M.}\ \bibnamefont {Fuller}}, \bibinfo {author} {\bibfnamefont {A.}~\bibnamefont {Kusenko}},\ and\ \bibinfo {author} {\bibfnamefont {V.}~\bibnamefont {Takhistov}},\ }\bibfield  {title} {\bibinfo {title} {{Primordial Black Holes and $r$-Process Nucleosynthesis}},\ }\href {https://doi.org/10.1103/PhysRevLett.119.061101} {\bibfield  {journal} {\bibinfo  {journal} {Phys. Rev. Lett.}\ }\textbf {\bibinfo {volume} {119}},\ \bibinfo {pages} {061101} (\bibinfo {year} {2017})},\ \Eprint {https://arxiv.org/abs/1704.01129} {arXiv:1704.01129 [astro-ph.HE]} \BibitemShut {NoStop}%
\bibitem [{\citenamefont {Takhistov}(2018)}]{Takhistov:2017bpt}%
  \BibitemOpen
  \bibfield  {author} {\bibinfo {author} {\bibfnamefont {V.}~\bibnamefont {Takhistov}},\ }\bibfield  {title} {\bibinfo {title} {{Transmuted Gravity Wave Signals from Primordial Black Holes}},\ }\href {https://doi.org/10.1016/j.physletb.2018.05.026} {\bibfield  {journal} {\bibinfo  {journal} {Phys. Lett. B}\ }\textbf {\bibinfo {volume} {782}},\ \bibinfo {pages} {77} (\bibinfo {year} {2018})},\ \Eprint {https://arxiv.org/abs/1707.05849} {arXiv:1707.05849 [astro-ph.CO]} \BibitemShut {NoStop}%
\bibitem [{\citenamefont {Takhistov}(2019)}]{Takhistov:2017nmt}%
  \BibitemOpen
  \bibfield  {author} {\bibinfo {author} {\bibfnamefont {V.}~\bibnamefont {Takhistov}},\ }\bibfield  {title} {\bibinfo {title} {{Positrons from Primordial Black Hole Microquasars and Gamma-ray Bursts}},\ }\href {https://doi.org/10.1016/j.physletb.2018.12.043} {\bibfield  {journal} {\bibinfo  {journal} {Phys. Lett. B}\ }\textbf {\bibinfo {volume} {789}},\ \bibinfo {pages} {538} (\bibinfo {year} {2019})},\ \Eprint {https://arxiv.org/abs/1710.09458} {arXiv:1710.09458 [astro-ph.HE]} \BibitemShut {NoStop}%
\bibitem [{\citenamefont {Bramante}\ \emph {et~al.}(2018)\citenamefont {Bramante}, \citenamefont {Linden},\ and\ \citenamefont {Tsai}}]{Bramante:2017ulk}%
  \BibitemOpen
  \bibfield  {author} {\bibinfo {author} {\bibfnamefont {J.}~\bibnamefont {Bramante}}, \bibinfo {author} {\bibfnamefont {T.}~\bibnamefont {Linden}},\ and\ \bibinfo {author} {\bibfnamefont {Y.-D.}\ \bibnamefont {Tsai}},\ }\bibfield  {title} {\bibinfo {title} {{Searching for dark matter with neutron star mergers and quiet kilonovae}},\ }\href {https://doi.org/10.1103/PhysRevD.97.055016} {\bibfield  {journal} {\bibinfo  {journal} {Phys. Rev. D}\ }\textbf {\bibinfo {volume} {97}},\ \bibinfo {pages} {055016} (\bibinfo {year} {2018})},\ \Eprint {https://arxiv.org/abs/1706.00001} {arXiv:1706.00001 [hep-ph]} \BibitemShut {NoStop}%
\bibitem [{\citenamefont {Takhistov}\ \emph {et~al.}(2021)\citenamefont {Takhistov}, \citenamefont {Fuller},\ and\ \citenamefont {Kusenko}}]{Takhistov:2020vxs}%
  \BibitemOpen
  \bibfield  {author} {\bibinfo {author} {\bibfnamefont {V.}~\bibnamefont {Takhistov}}, \bibinfo {author} {\bibfnamefont {G.~M.}\ \bibnamefont {Fuller}},\ and\ \bibinfo {author} {\bibfnamefont {A.}~\bibnamefont {Kusenko}},\ }\bibfield  {title} {\bibinfo {title} {{Test for the Origin of Solar Mass Black Holes}},\ }\href {https://doi.org/10.1103/PhysRevLett.126.071101} {\bibfield  {journal} {\bibinfo  {journal} {Phys. Rev. Lett.}\ }\textbf {\bibinfo {volume} {126}},\ \bibinfo {pages} {071101} (\bibinfo {year} {2021})},\ \Eprint {https://arxiv.org/abs/2008.12780} {arXiv:2008.12780 [astro-ph.HE]} \BibitemShut {NoStop}%
\bibitem [{\citenamefont {Dasgupta}\ \emph {et~al.}(2021)\citenamefont {Dasgupta}, \citenamefont {Laha},\ and\ \citenamefont {Ray}}]{Dasgupta:2020mqg}%
  \BibitemOpen
  \bibfield  {author} {\bibinfo {author} {\bibfnamefont {B.}~\bibnamefont {Dasgupta}}, \bibinfo {author} {\bibfnamefont {R.}~\bibnamefont {Laha}},\ and\ \bibinfo {author} {\bibfnamefont {A.}~\bibnamefont {Ray}},\ }\bibfield  {title} {\bibinfo {title} {{Low Mass Black Holes from Dark Core Collapse}},\ }\href {https://doi.org/10.1103/PhysRevLett.126.141105} {\bibfield  {journal} {\bibinfo  {journal} {Phys. Rev. Lett.}\ }\textbf {\bibinfo {volume} {126}},\ \bibinfo {pages} {141105} (\bibinfo {year} {2021})},\ \Eprint {https://arxiv.org/abs/2009.01825} {arXiv:2009.01825 [astro-ph.HE]} \BibitemShut {NoStop}%
\bibitem [{\citenamefont {Wang}\ and\ \citenamefont {Zhao}(2022)}]{Wang:2021iwp}%
  \BibitemOpen
  \bibfield  {author} {\bibinfo {author} {\bibfnamefont {S.}~\bibnamefont {Wang}}\ and\ \bibinfo {author} {\bibfnamefont {Z.-C.}\ \bibnamefont {Zhao}},\ }\bibfield  {title} {\bibinfo {title} {{GW200105 and GW200115 are compatible with a scenario of primordial black hole binary coalescences}},\ }\href {https://doi.org/10.1140/epjc/s10052-021-09981-1} {\bibfield  {journal} {\bibinfo  {journal} {Eur. Phys. J. C}\ }\textbf {\bibinfo {volume} {82}},\ \bibinfo {pages} {9} (\bibinfo {year} {2022})},\ \Eprint {https://arxiv.org/abs/2107.00450} {arXiv:2107.00450 [astro-ph.CO]} \BibitemShut {NoStop}%
\bibitem [{\citenamefont {Sasaki}\ \emph {et~al.}(2022)\citenamefont {Sasaki}, \citenamefont {Takhistov}, \citenamefont {Vardanyan},\ and\ \citenamefont {Zhang}}]{Sasaki:2021iuc}%
  \BibitemOpen
  \bibfield  {author} {\bibinfo {author} {\bibfnamefont {M.}~\bibnamefont {Sasaki}}, \bibinfo {author} {\bibfnamefont {V.}~\bibnamefont {Takhistov}}, \bibinfo {author} {\bibfnamefont {V.}~\bibnamefont {Vardanyan}},\ and\ \bibinfo {author} {\bibfnamefont {Y.-l.}\ \bibnamefont {Zhang}},\ }\bibfield  {title} {\bibinfo {title} {{Establishing the Nonprimordial Origin of Black Hole\textendash{}Neutron Star Mergers}},\ }\href {https://doi.org/10.3847/1538-4357/ac66da} {\bibfield  {journal} {\bibinfo  {journal} {Astrophys. J.}\ }\textbf {\bibinfo {volume} {931}},\ \bibinfo {pages} {2} (\bibinfo {year} {2022})},\ \Eprint {https://arxiv.org/abs/2110.09509} {arXiv:2110.09509 [astro-ph.CO]} \BibitemShut {NoStop}%
\bibitem [{\citenamefont {Abbott}\ \emph {et~al.}(2022)\citenamefont {Abbott} \emph {et~al.}}]{LIGOScientific:2022hai}%
  \BibitemOpen
  \bibfield  {author} {\bibinfo {author} {\bibfnamefont {R.}~\bibnamefont {Abbott}} \emph {et~al.} (\bibinfo {collaboration} {LIGO Scientific, VIRGO, KAGRA}),\ }\bibfield  {title} {\bibinfo {title} {{Search for subsolar-mass black hole binaries in the second part of Advanced LIGO's and Advanced Virgo's third observing run}},\ }\href@noop {} {\  (\bibinfo {year} {2022})},\ \Eprint {https://arxiv.org/abs/2212.01477} {arXiv:2212.01477 [astro-ph.HE]} \BibitemShut {NoStop}%
\bibitem [{\citenamefont {Bertschinger}(1985)}]{Bertschinger:1985pd}%
  \BibitemOpen
  \bibfield  {author} {\bibinfo {author} {\bibfnamefont {E.}~\bibnamefont {Bertschinger}},\ }\bibfield  {title} {\bibinfo {title} {{Self - similar secondary infall and accretion in an Einstein-de Sitter universe}},\ }\href {https://doi.org/10.1086/191028} {\bibfield  {journal} {\bibinfo  {journal} {Astrophys. J. Suppl.}\ }\textbf {\bibinfo {volume} {58}},\ \bibinfo {pages} {39} (\bibinfo {year} {1985})}\BibitemShut {NoStop}%
\bibitem [{\citenamefont {Mack}\ \emph {et~al.}(2007)\citenamefont {Mack}, \citenamefont {Ostriker},\ and\ \citenamefont {Ricotti}}]{Mack:2006gz}%
  \BibitemOpen
  \bibfield  {author} {\bibinfo {author} {\bibfnamefont {K.~J.}\ \bibnamefont {Mack}}, \bibinfo {author} {\bibfnamefont {J.~P.}\ \bibnamefont {Ostriker}},\ and\ \bibinfo {author} {\bibfnamefont {M.}~\bibnamefont {Ricotti}},\ }\bibfield  {title} {\bibinfo {title} {{Growth of structure seeded by primordial black holes}},\ }\href {https://doi.org/10.1086/518998} {\bibfield  {journal} {\bibinfo  {journal} {Astrophys. J.}\ }\textbf {\bibinfo {volume} {665}},\ \bibinfo {pages} {1277} (\bibinfo {year} {2007})},\ \Eprint {https://arxiv.org/abs/astro-ph/0608642} {arXiv:astro-ph/0608642} \BibitemShut {NoStop}%
\bibitem [{\citenamefont {Ricotti}\ \emph {et~al.}(2008)\citenamefont {Ricotti}, \citenamefont {Ostriker},\ and\ \citenamefont {Mack}}]{Ricotti:2007au}%
  \BibitemOpen
  \bibfield  {author} {\bibinfo {author} {\bibfnamefont {M.}~\bibnamefont {Ricotti}}, \bibinfo {author} {\bibfnamefont {J.~P.}\ \bibnamefont {Ostriker}},\ and\ \bibinfo {author} {\bibfnamefont {K.~J.}\ \bibnamefont {Mack}},\ }\bibfield  {title} {\bibinfo {title} {{Effect of Primordial Black Holes on the Cosmic Microwave Background and Cosmological Parameter Estimates}},\ }\href {https://doi.org/10.1086/587831} {\bibfield  {journal} {\bibinfo  {journal} {Astrophys. J.}\ }\textbf {\bibinfo {volume} {680}},\ \bibinfo {pages} {829} (\bibinfo {year} {2008})},\ \Eprint {https://arxiv.org/abs/0709.0524} {arXiv:0709.0524 [astro-ph]} \BibitemShut {NoStop}%
\bibitem [{\citenamefont {Zhao}\ and\ \citenamefont {Silk}(2005)}]{Zhao:2005zr}%
  \BibitemOpen
  \bibfield  {author} {\bibinfo {author} {\bibfnamefont {H.-S.}\ \bibnamefont {Zhao}}\ and\ \bibinfo {author} {\bibfnamefont {J.}~\bibnamefont {Silk}},\ }\bibfield  {title} {\bibinfo {title} {{Mini-dark halos with intermediate mass black holes}},\ }\href {https://doi.org/10.1103/PhysRevLett.95.011301} {\bibfield  {journal} {\bibinfo  {journal} {Phys. Rev. Lett.}\ }\textbf {\bibinfo {volume} {95}},\ \bibinfo {pages} {011301} (\bibinfo {year} {2005})},\ \Eprint {https://arxiv.org/abs/astro-ph/0501625} {arXiv:astro-ph/0501625} \BibitemShut {NoStop}%
\bibitem [{\citenamefont {Bertone}\ \emph {et~al.}(2005)\citenamefont {Bertone}, \citenamefont {Zentner},\ and\ \citenamefont {Silk}}]{Bertone:2005xz}%
  \BibitemOpen
  \bibfield  {author} {\bibinfo {author} {\bibfnamefont {G.}~\bibnamefont {Bertone}}, \bibinfo {author} {\bibfnamefont {A.~R.}\ \bibnamefont {Zentner}},\ and\ \bibinfo {author} {\bibfnamefont {J.}~\bibnamefont {Silk}},\ }\bibfield  {title} {\bibinfo {title} {{A new signature of dark matter annihilations: gamma-rays from intermediate-mass black holes}},\ }\href {https://doi.org/10.1103/PhysRevD.72.103517} {\bibfield  {journal} {\bibinfo  {journal} {Phys. Rev. D}\ }\textbf {\bibinfo {volume} {72}},\ \bibinfo {pages} {103517} (\bibinfo {year} {2005})},\ \Eprint {https://arxiv.org/abs/astro-ph/0509565} {arXiv:astro-ph/0509565} \BibitemShut {NoStop}%
\bibitem [{\citenamefont {Bringmann}\ \emph {et~al.}(2009)\citenamefont {Bringmann}, \citenamefont {Lavalle},\ and\ \citenamefont {Salati}}]{Bringmann:2009}%
  \BibitemOpen
  \bibfield  {author} {\bibinfo {author} {\bibfnamefont {T.}~\bibnamefont {Bringmann}}, \bibinfo {author} {\bibfnamefont {J.}~\bibnamefont {Lavalle}},\ and\ \bibinfo {author} {\bibfnamefont {P.}~\bibnamefont {Salati}},\ }\bibfield  {title} {\bibinfo {title} {Intermediate mass black holes and nearby dark matter point sources: A critical reassessment},\ }\bibfield  {journal} {\bibinfo  {journal} {Physical Review Letters}\ }\textbf {\bibinfo {volume} {103}},\ \href {https://doi.org/10.1103/physrevlett.103.161301} {10.1103/physrevlett.103.161301} (\bibinfo {year} {2009})\BibitemShut {NoStop}%
\bibitem [{\citenamefont {Aschersleben}\ \emph {et~al.}(2024)\citenamefont {Aschersleben}, \citenamefont {Bertone}, \citenamefont {Horns}, \citenamefont {Moulin}, \citenamefont {Peletier},\ and\ \citenamefont {Vecchi}}]{Aschersleben:2024xsb}%
  \BibitemOpen
  \bibfield  {author} {\bibinfo {author} {\bibfnamefont {J.}~\bibnamefont {Aschersleben}}, \bibinfo {author} {\bibfnamefont {G.}~\bibnamefont {Bertone}}, \bibinfo {author} {\bibfnamefont {D.}~\bibnamefont {Horns}}, \bibinfo {author} {\bibfnamefont {E.}~\bibnamefont {Moulin}}, \bibinfo {author} {\bibfnamefont {R.~F.}\ \bibnamefont {Peletier}},\ and\ \bibinfo {author} {\bibfnamefont {M.}~\bibnamefont {Vecchi}},\ }\bibfield  {title} {\bibinfo {title} {{Gamma rays from dark matter spikes in EAGLE simulations}},\ }\href@noop {} {\  (\bibinfo {year} {2024})},\ \Eprint {https://arxiv.org/abs/2401.14072} {arXiv:2401.14072 [astro-ph.HE]} \BibitemShut {NoStop}%
\bibitem [{\citenamefont {Bertone}\ \emph {et~al.}(2024)\citenamefont {Bertone}, \citenamefont {Wierda}, \citenamefont {Gaggero}, \citenamefont {Kavanagh}, \citenamefont {Volonteri},\ and\ \citenamefont {Yoshida}}]{Bertone:2024wbn}%
  \BibitemOpen
  \bibfield  {author} {\bibinfo {author} {\bibfnamefont {G.}~\bibnamefont {Bertone}}, \bibinfo {author} {\bibfnamefont {A.~R. A.~C.}\ \bibnamefont {Wierda}}, \bibinfo {author} {\bibfnamefont {D.}~\bibnamefont {Gaggero}}, \bibinfo {author} {\bibfnamefont {B.~J.}\ \bibnamefont {Kavanagh}}, \bibinfo {author} {\bibfnamefont {M.}~\bibnamefont {Volonteri}},\ and\ \bibinfo {author} {\bibfnamefont {N.}~\bibnamefont {Yoshida}},\ }\bibfield  {title} {\bibinfo {title} {{Dark Matter Mounds: towards a realistic description of dark matter overdensities around black holes}},\ }\href@noop {} {\  (\bibinfo {year} {2024})},\ \Eprint {https://arxiv.org/abs/2404.08731} {arXiv:2404.08731 [astro-ph.CO]} \BibitemShut {NoStop}%
\bibitem [{\citenamefont {Adamek}\ \emph {et~al.}(2019)\citenamefont {Adamek}, \citenamefont {Byrnes}, \citenamefont {Gosenca},\ and\ \citenamefont {Hotchkiss}}]{Adamek:2019gns}%
  \BibitemOpen
  \bibfield  {author} {\bibinfo {author} {\bibfnamefont {J.}~\bibnamefont {Adamek}}, \bibinfo {author} {\bibfnamefont {C.~T.}\ \bibnamefont {Byrnes}}, \bibinfo {author} {\bibfnamefont {M.}~\bibnamefont {Gosenca}},\ and\ \bibinfo {author} {\bibfnamefont {S.}~\bibnamefont {Hotchkiss}},\ }\bibfield  {title} {\bibinfo {title} {{WIMPs and stellar-mass primordial black holes are incompatible}},\ }\href {https://doi.org/10.1103/PhysRevD.100.023506} {\bibfield  {journal} {\bibinfo  {journal} {Phys. Rev. D}\ }\textbf {\bibinfo {volume} {100}},\ \bibinfo {pages} {023506} (\bibinfo {year} {2019})},\ \Eprint {https://arxiv.org/abs/1901.08528} {arXiv:1901.08528 [astro-ph.CO]} \BibitemShut {NoStop}%
\bibitem [{\citenamefont {Lacki}\ and\ \citenamefont {Beacom}(2010)}]{Lacki:2010zf}%
  \BibitemOpen
  \bibfield  {author} {\bibinfo {author} {\bibfnamefont {B.~C.}\ \bibnamefont {Lacki}}\ and\ \bibinfo {author} {\bibfnamefont {J.~F.}\ \bibnamefont {Beacom}},\ }\bibfield  {title} {\bibinfo {title} {{Primordial Black Holes as Dark Matter: Almost All or Almost Nothing}},\ }\href {https://doi.org/10.1088/2041-8205/720/1/L67} {\bibfield  {journal} {\bibinfo  {journal} {Astrophys. J. Lett.}\ }\textbf {\bibinfo {volume} {720}},\ \bibinfo {pages} {L67} (\bibinfo {year} {2010})},\ \Eprint {https://arxiv.org/abs/1003.3466} {arXiv:1003.3466 [astro-ph.CO]} \BibitemShut {NoStop}%
\bibitem [{\citenamefont {Hertzberg}\ \emph {et~al.}(2021)\citenamefont {Hertzberg}, \citenamefont {Nurmi}, \citenamefont {Schiappacasse},\ and\ \citenamefont {Yanagida}}]{Hertzberg:2020kpm}%
  \BibitemOpen
  \bibfield  {author} {\bibinfo {author} {\bibfnamefont {M.~P.}\ \bibnamefont {Hertzberg}}, \bibinfo {author} {\bibfnamefont {S.}~\bibnamefont {Nurmi}}, \bibinfo {author} {\bibfnamefont {E.~D.}\ \bibnamefont {Schiappacasse}},\ and\ \bibinfo {author} {\bibfnamefont {T.~T.}\ \bibnamefont {Yanagida}},\ }\bibfield  {title} {\bibinfo {title} {{Shining Primordial Black Holes}},\ }\href {https://doi.org/10.1103/PhysRevD.103.063025} {\bibfield  {journal} {\bibinfo  {journal} {Phys. Rev. D}\ }\textbf {\bibinfo {volume} {103}},\ \bibinfo {pages} {063025} (\bibinfo {year} {2021})},\ \Eprint {https://arxiv.org/abs/2011.05922} {arXiv:2011.05922 [hep-ph]} \BibitemShut {NoStop}%
\bibitem [{\citenamefont {Nurmi}\ \emph {et~al.}(2021)\citenamefont {Nurmi}, \citenamefont {Schiappacasse},\ and\ \citenamefont {Yanagida}}]{Nurmi:2021xds}%
  \BibitemOpen
  \bibfield  {author} {\bibinfo {author} {\bibfnamefont {S.}~\bibnamefont {Nurmi}}, \bibinfo {author} {\bibfnamefont {E.~D.}\ \bibnamefont {Schiappacasse}},\ and\ \bibinfo {author} {\bibfnamefont {T.~T.}\ \bibnamefont {Yanagida}},\ }\bibfield  {title} {\bibinfo {title} {{Radio signatures from encounters between neutron stars and QCD-axion minihalos around primordial black~holes}},\ }\href {https://doi.org/10.1088/1475-7516/2021/09/004} {\bibfield  {journal} {\bibinfo  {journal} {JCAP}\ }\textbf {\bibinfo {volume} {09}},\ \bibinfo {pages} {004}},\ \Eprint {https://arxiv.org/abs/2102.05680} {arXiv:2102.05680 [hep-ph]} \BibitemShut {NoStop}%
\bibitem [{\citenamefont {Oguri}\ \emph {et~al.}(2023)\citenamefont {Oguri}, \citenamefont {Takhistov},\ and\ \citenamefont {Kohri}}]{Oguri:2022fir}%
  \BibitemOpen
  \bibfield  {author} {\bibinfo {author} {\bibfnamefont {M.}~\bibnamefont {Oguri}}, \bibinfo {author} {\bibfnamefont {V.}~\bibnamefont {Takhistov}},\ and\ \bibinfo {author} {\bibfnamefont {K.}~\bibnamefont {Kohri}},\ }\bibfield  {title} {\bibinfo {title} {{Revealing dark matter dress of primordial black holes by cosmological lensing}},\ }\href {https://doi.org/10.1016/j.physletb.2023.138276} {\bibfield  {journal} {\bibinfo  {journal} {Phys. Lett. B}\ }\textbf {\bibinfo {volume} {847}},\ \bibinfo {pages} {138276} (\bibinfo {year} {2023})},\ \Eprint {https://arxiv.org/abs/2208.05957} {arXiv:2208.05957 [astro-ph.CO]} \BibitemShut {NoStop}%
\bibitem [{\citenamefont {Cai}\ \emph {et~al.}(2023)\citenamefont {Cai}, \citenamefont {Chen}, \citenamefont {Wang},\ and\ \citenamefont {Yang}}]{Cai:2022kbp}%
  \BibitemOpen
  \bibfield  {author} {\bibinfo {author} {\bibfnamefont {R.-G.}\ \bibnamefont {Cai}}, \bibinfo {author} {\bibfnamefont {T.}~\bibnamefont {Chen}}, \bibinfo {author} {\bibfnamefont {S.-J.}\ \bibnamefont {Wang}},\ and\ \bibinfo {author} {\bibfnamefont {X.-Y.}\ \bibnamefont {Yang}},\ }\bibfield  {title} {\bibinfo {title} {{Gravitational microlensing by dressed primordial black holes}},\ }\href {https://doi.org/10.1088/1475-7516/2023/03/043} {\bibfield  {journal} {\bibinfo  {journal} {JCAP}\ }\textbf {\bibinfo {volume} {03}},\ \bibinfo {pages} {043}},\ \Eprint {https://arxiv.org/abs/2210.02078} {arXiv:2210.02078 [astro-ph.CO]} \BibitemShut {NoStop}%
\bibitem [{\citenamefont {Eda}\ \emph {et~al.}(2015)\citenamefont {Eda}, \citenamefont {Itoh}, \citenamefont {Kuroyanagi},\ and\ \citenamefont {Silk}}]{Eda:2014kra}%
  \BibitemOpen
  \bibfield  {author} {\bibinfo {author} {\bibfnamefont {K.}~\bibnamefont {Eda}}, \bibinfo {author} {\bibfnamefont {Y.}~\bibnamefont {Itoh}}, \bibinfo {author} {\bibfnamefont {S.}~\bibnamefont {Kuroyanagi}},\ and\ \bibinfo {author} {\bibfnamefont {J.}~\bibnamefont {Silk}},\ }\bibfield  {title} {\bibinfo {title} {{Gravitational waves as a probe of dark matter minispikes}},\ }\href {https://doi.org/10.1103/PhysRevD.91.044045} {\bibfield  {journal} {\bibinfo  {journal} {Phys. Rev. D}\ }\textbf {\bibinfo {volume} {91}},\ \bibinfo {pages} {044045} (\bibinfo {year} {2015})},\ \Eprint {https://arxiv.org/abs/1408.3534} {arXiv:1408.3534 [gr-qc]} \BibitemShut {NoStop}%
\bibitem [{\citenamefont {Kavanagh}\ \emph {et~al.}(2020)\citenamefont {Kavanagh}, \citenamefont {Nichols}, \citenamefont {Bertone},\ and\ \citenamefont {Gaggero}}]{Kavanagh:2020cfn}%
  \BibitemOpen
  \bibfield  {author} {\bibinfo {author} {\bibfnamefont {B.~J.}\ \bibnamefont {Kavanagh}}, \bibinfo {author} {\bibfnamefont {D.~A.}\ \bibnamefont {Nichols}}, \bibinfo {author} {\bibfnamefont {G.}~\bibnamefont {Bertone}},\ and\ \bibinfo {author} {\bibfnamefont {D.}~\bibnamefont {Gaggero}},\ }\bibfield  {title} {\bibinfo {title} {{Detecting dark matter around black holes with gravitational waves: Effects of dark-matter dynamics on the gravitational waveform}},\ }\href {https://doi.org/10.1103/PhysRevD.102.083006} {\bibfield  {journal} {\bibinfo  {journal} {Phys. Rev. D}\ }\textbf {\bibinfo {volume} {102}},\ \bibinfo {pages} {083006} (\bibinfo {year} {2020})},\ \Eprint {https://arxiv.org/abs/2002.12811} {arXiv:2002.12811 [gr-qc]} \BibitemShut {NoStop}%
\bibitem [{\citenamefont {Coogan}\ \emph {et~al.}(2022)\citenamefont {Coogan}, \citenamefont {Bertone}, \citenamefont {Gaggero}, \citenamefont {Kavanagh},\ and\ \citenamefont {Nichols}}]{Coogan:2021uqv}%
  \BibitemOpen
  \bibfield  {author} {\bibinfo {author} {\bibfnamefont {A.}~\bibnamefont {Coogan}}, \bibinfo {author} {\bibfnamefont {G.}~\bibnamefont {Bertone}}, \bibinfo {author} {\bibfnamefont {D.}~\bibnamefont {Gaggero}}, \bibinfo {author} {\bibfnamefont {B.~J.}\ \bibnamefont {Kavanagh}},\ and\ \bibinfo {author} {\bibfnamefont {D.~A.}\ \bibnamefont {Nichols}},\ }\bibfield  {title} {\bibinfo {title} {{Measuring the dark matter environments of black hole binaries with gravitational waves}},\ }\href {https://doi.org/10.1103/PhysRevD.105.043009} {\bibfield  {journal} {\bibinfo  {journal} {Phys. Rev. D}\ }\textbf {\bibinfo {volume} {105}},\ \bibinfo {pages} {043009} (\bibinfo {year} {2022})},\ \Eprint {https://arxiv.org/abs/2108.04154} {arXiv:2108.04154 [gr-qc]} \BibitemShut {NoStop}%
\bibitem [{\citenamefont {Cole}\ \emph {et~al.}(2023{\natexlab{a}})\citenamefont {Cole}, \citenamefont {Coogan}, \citenamefont {Kavanagh},\ and\ \citenamefont {Bertone}}]{Cole:2022ucw}%
  \BibitemOpen
  \bibfield  {author} {\bibinfo {author} {\bibfnamefont {P.~S.}\ \bibnamefont {Cole}}, \bibinfo {author} {\bibfnamefont {A.}~\bibnamefont {Coogan}}, \bibinfo {author} {\bibfnamefont {B.~J.}\ \bibnamefont {Kavanagh}},\ and\ \bibinfo {author} {\bibfnamefont {G.}~\bibnamefont {Bertone}},\ }\bibfield  {title} {\bibinfo {title} {{Measuring dark matter spikes around primordial black holes with Einstein Telescope and Cosmic Explorer}},\ }\href {https://doi.org/10.1103/PhysRevD.107.083006} {\bibfield  {journal} {\bibinfo  {journal} {Phys. Rev. D}\ }\textbf {\bibinfo {volume} {107}},\ \bibinfo {pages} {083006} (\bibinfo {year} {2023}{\natexlab{a}})},\ \Eprint {https://arxiv.org/abs/2207.07576} {arXiv:2207.07576 [astro-ph.CO]} \BibitemShut {NoStop}%
\bibitem [{\citenamefont {Cole}\ \emph {et~al.}(2023{\natexlab{b}})\citenamefont {Cole}, \citenamefont {Bertone}, \citenamefont {Coogan}, \citenamefont {Gaggero}, \citenamefont {Karydas}, \citenamefont {Kavanagh}, \citenamefont {Spieksma},\ and\ \citenamefont {Tomaselli}}]{Cole:2022yzw}%
  \BibitemOpen
  \bibfield  {author} {\bibinfo {author} {\bibfnamefont {P.~S.}\ \bibnamefont {Cole}}, \bibinfo {author} {\bibfnamefont {G.}~\bibnamefont {Bertone}}, \bibinfo {author} {\bibfnamefont {A.}~\bibnamefont {Coogan}}, \bibinfo {author} {\bibfnamefont {D.}~\bibnamefont {Gaggero}}, \bibinfo {author} {\bibfnamefont {T.}~\bibnamefont {Karydas}}, \bibinfo {author} {\bibfnamefont {B.~J.}\ \bibnamefont {Kavanagh}}, \bibinfo {author} {\bibfnamefont {T.~F.~M.}\ \bibnamefont {Spieksma}},\ and\ \bibinfo {author} {\bibfnamefont {G.~M.}\ \bibnamefont {Tomaselli}},\ }\bibfield  {title} {\bibinfo {title} {{Distinguishing environmental effects on binary black hole gravitational waveforms}},\ }\href {https://doi.org/10.1038/s41550-023-01990-2} {\bibfield  {journal} {\bibinfo  {journal} {Nature Astron.}\ }\textbf {\bibinfo {volume} {7}},\ \bibinfo {pages} {943} (\bibinfo {year} {2023}{\natexlab{b}})},\ \Eprint {https://arxiv.org/abs/2211.01362} {arXiv:2211.01362 [gr-qc]} \BibitemShut {NoStop}%
\bibitem [{\citenamefont {Nakamura}(1998)}]{Nakamura:1997sw}%
  \BibitemOpen
  \bibfield  {author} {\bibinfo {author} {\bibfnamefont {T.~T.}\ \bibnamefont {Nakamura}},\ }\bibfield  {title} {\bibinfo {title} {{Gravitational lensing of gravitational waves from inspiraling binaries by a point mass lens}},\ }\href {https://doi.org/10.1103/PhysRevLett.80.1138} {\bibfield  {journal} {\bibinfo  {journal} {Phys. Rev. Lett.}\ }\textbf {\bibinfo {volume} {80}},\ \bibinfo {pages} {1138} (\bibinfo {year} {1998})}\BibitemShut {NoStop}%
\bibitem [{\citenamefont {Takahashi}\ and\ \citenamefont {Nakamura}(2003)}]{takahashi2003wave}%
  \BibitemOpen
  \bibfield  {author} {\bibinfo {author} {\bibfnamefont {R.}~\bibnamefont {Takahashi}}\ and\ \bibinfo {author} {\bibfnamefont {T.}~\bibnamefont {Nakamura}},\ }\bibfield  {title} {\bibinfo {title} {Wave effects in the gravitational lensing of gravitational waves from chirping binaries},\ }\href@noop {} {\bibfield  {journal} {\bibinfo  {journal} {The Astrophysical Journal}\ }\textbf {\bibinfo {volume} {595}},\ \bibinfo {pages} {1039} (\bibinfo {year} {2003})}\BibitemShut {NoStop}%
\bibitem [{\citenamefont {Jung}\ and\ \citenamefont {Shin}(2019)}]{Jung:2017flg}%
  \BibitemOpen
  \bibfield  {author} {\bibinfo {author} {\bibfnamefont {S.}~\bibnamefont {Jung}}\ and\ \bibinfo {author} {\bibfnamefont {C.~S.}\ \bibnamefont {Shin}},\ }\bibfield  {title} {\bibinfo {title} {{Gravitational-Wave Fringes at LIGO: Detecting Compact Dark Matter by Gravitational Lensing}},\ }\href {https://doi.org/10.1103/PhysRevLett.122.041103} {\bibfield  {journal} {\bibinfo  {journal} {Phys. Rev. Lett.}\ }\textbf {\bibinfo {volume} {122}},\ \bibinfo {pages} {041103} (\bibinfo {year} {2019})},\ \Eprint {https://arxiv.org/abs/1712.01396} {arXiv:1712.01396 [astro-ph.CO]} \BibitemShut {NoStop}%
\bibitem [{\citenamefont {Lai}\ \emph {et~al.}(2018)\citenamefont {Lai}, \citenamefont {Hannuksela}, \citenamefont {Herrera-Mart{\'\i}n}, \citenamefont {Diego}, \citenamefont {Broadhurst},\ and\ \citenamefont {Li}}]{lai2018discovering}%
  \BibitemOpen
  \bibfield  {author} {\bibinfo {author} {\bibfnamefont {K.-H.}\ \bibnamefont {Lai}}, \bibinfo {author} {\bibfnamefont {O.~A.}\ \bibnamefont {Hannuksela}}, \bibinfo {author} {\bibfnamefont {A.}~\bibnamefont {Herrera-Mart{\'\i}n}}, \bibinfo {author} {\bibfnamefont {J.~M.}\ \bibnamefont {Diego}}, \bibinfo {author} {\bibfnamefont {T.}~\bibnamefont {Broadhurst}},\ and\ \bibinfo {author} {\bibfnamefont {T.~G.}\ \bibnamefont {Li}},\ }\bibfield  {title} {\bibinfo {title} {Discovering intermediate-mass black hole lenses through gravitational wave lensing},\ }\href@noop {} {\bibfield  {journal} {\bibinfo  {journal} {Physical Review D}\ }\textbf {\bibinfo {volume} {98}},\ \bibinfo {pages} {083005} (\bibinfo {year} {2018})}\BibitemShut {NoStop}%
\bibitem [{\citenamefont {Urrutia}\ and\ \citenamefont {Vaskonen}(2021)}]{Urrutia:2021qak}%
  \BibitemOpen
  \bibfield  {author} {\bibinfo {author} {\bibfnamefont {J.}~\bibnamefont {Urrutia}}\ and\ \bibinfo {author} {\bibfnamefont {V.}~\bibnamefont {Vaskonen}},\ }\bibfield  {title} {\bibinfo {title} {{Lensing of gravitational waves as a probe of compact dark matter}},\ }\href {https://doi.org/10.1093/mnras/stab3118} {\bibfield  {journal} {\bibinfo  {journal} {Mon. Not. Roy. Astron. Soc.}\ }\textbf {\bibinfo {volume} {509}},\ \bibinfo {pages} {1358} (\bibinfo {year} {2021})},\ \Eprint {https://arxiv.org/abs/2109.03213} {arXiv:2109.03213 [astro-ph.CO]} \BibitemShut {NoStop}%
\bibitem [{\citenamefont {Basak}\ \emph {et~al.}(2022)\citenamefont {Basak}, \citenamefont {Ganguly}, \citenamefont {Haris}, \citenamefont {Kapadia}, \citenamefont {Mehta},\ and\ \citenamefont {Ajith}}]{Basak:2021ten}%
  \BibitemOpen
  \bibfield  {author} {\bibinfo {author} {\bibfnamefont {S.}~\bibnamefont {Basak}}, \bibinfo {author} {\bibfnamefont {A.}~\bibnamefont {Ganguly}}, \bibinfo {author} {\bibfnamefont {K.}~\bibnamefont {Haris}}, \bibinfo {author} {\bibfnamefont {S.}~\bibnamefont {Kapadia}}, \bibinfo {author} {\bibfnamefont {A.~K.}\ \bibnamefont {Mehta}},\ and\ \bibinfo {author} {\bibfnamefont {P.}~\bibnamefont {Ajith}},\ }\bibfield  {title} {\bibinfo {title} {{Constraints on Compact Dark Matter from Gravitational Wave Microlensing}},\ }\href {https://doi.org/10.3847/2041-8213/ac4dfa} {\bibfield  {journal} {\bibinfo  {journal} {Astrophys. J.}\ }\textbf {\bibinfo {volume} {926}},\ \bibinfo {pages} {L28} (\bibinfo {year} {2022})},\ \Eprint {https://arxiv.org/abs/2109.06456} {arXiv:2109.06456 [gr-qc]} \BibitemShut {NoStop}%
\bibitem [{\citenamefont {Abbott}\ \emph {et~al.}(2021{\natexlab{a}})\citenamefont {Abbott} \emph {et~al.}}]{LIGOScientific:2021izm}%
  \BibitemOpen
  \bibfield  {author} {\bibinfo {author} {\bibfnamefont {R.}~\bibnamefont {Abbott}} \emph {et~al.} (\bibinfo {collaboration} {LIGO Scientific, VIRGO}),\ }\bibfield  {title} {\bibinfo {title} {{Search for Lensing Signatures in the Gravitational-Wave Observations from the First Half of LIGO\textendash{}Virgo\textquoteright{}s Third Observing Run}},\ }\href {https://doi.org/10.3847/1538-4357/ac23db} {\bibfield  {journal} {\bibinfo  {journal} {Astrophys. J.}\ }\textbf {\bibinfo {volume} {923}},\ \bibinfo {pages} {14} (\bibinfo {year} {2021}{\natexlab{a}})},\ \Eprint {https://arxiv.org/abs/2105.06384} {arXiv:2105.06384 [gr-qc]} \BibitemShut {NoStop}%
\bibitem [{\citenamefont {Abbott}\ \emph {et~al.}(2023{\natexlab{a}})\citenamefont {Abbott} \emph {et~al.}}]{LIGOScientific:2023bwz}%
  \BibitemOpen
  \bibfield  {author} {\bibinfo {author} {\bibfnamefont {R.}~\bibnamefont {Abbott}} \emph {et~al.} (\bibinfo {collaboration} {LIGO Scientific, VIRGO, KAGRA}),\ }\bibfield  {title} {\bibinfo {title} {{Search for gravitational-lensing signatures in the full third observing run of the LIGO-Virgo network}},\ }\href@noop {} {\  (\bibinfo {year} {2023}{\natexlab{a}})},\ \Eprint {https://arxiv.org/abs/2304.08393} {arXiv:2304.08393 [gr-qc]} \BibitemShut {NoStop}%
\bibitem [{\citenamefont {Zhou}\ \emph {et~al.}(2022)\citenamefont {Zhou}, \citenamefont {Li}, \citenamefont {Liao},\ and\ \citenamefont {Huang}}]{Zhou:2022yeo}%
  \BibitemOpen
  \bibfield  {author} {\bibinfo {author} {\bibfnamefont {H.}~\bibnamefont {Zhou}}, \bibinfo {author} {\bibfnamefont {Z.}~\bibnamefont {Li}}, \bibinfo {author} {\bibfnamefont {K.}~\bibnamefont {Liao}},\ and\ \bibinfo {author} {\bibfnamefont {Z.}~\bibnamefont {Huang}},\ }\bibfield  {title} {\bibinfo {title} {{Constraints on compact dark matter from lensing of gravitational waves for the third-generation gravitational wave detector}},\ }\href {https://doi.org/10.1093/mnras/stac2944} {\bibfield  {journal} {\bibinfo  {journal} {Mon. Not. Roy. Astron. Soc.}\ }\textbf {\bibinfo {volume} {518}},\ \bibinfo {pages} {149} (\bibinfo {year} {2022})},\ \Eprint {https://arxiv.org/abs/2206.13128} {arXiv:2206.13128 [astro-ph.CO]} \BibitemShut {NoStop}%
\bibitem [{\citenamefont {Dai}\ \emph {et~al.}(2018)\citenamefont {Dai}, \citenamefont {Li}, \citenamefont {Zackay}, \citenamefont {Mao},\ and\ \citenamefont {Lu}}]{dai2018detecting}%
  \BibitemOpen
  \bibfield  {author} {\bibinfo {author} {\bibfnamefont {L.}~\bibnamefont {Dai}}, \bibinfo {author} {\bibfnamefont {S.-S.}\ \bibnamefont {Li}}, \bibinfo {author} {\bibfnamefont {B.}~\bibnamefont {Zackay}}, \bibinfo {author} {\bibfnamefont {S.}~\bibnamefont {Mao}},\ and\ \bibinfo {author} {\bibfnamefont {Y.}~\bibnamefont {Lu}},\ }\bibfield  {title} {\bibinfo {title} {Detecting lensing-induced diffraction in astrophysical gravitational waves},\ }\href@noop {} {\bibfield  {journal} {\bibinfo  {journal} {Physical Review D}\ }\textbf {\bibinfo {volume} {98}},\ \bibinfo {pages} {104029} (\bibinfo {year} {2018})}\BibitemShut {NoStop}%
\bibitem [{\citenamefont {Oguri}\ and\ \citenamefont {Takahashi}(2020)}]{oguri2020probing}%
  \BibitemOpen
  \bibfield  {author} {\bibinfo {author} {\bibfnamefont {M.}~\bibnamefont {Oguri}}\ and\ \bibinfo {author} {\bibfnamefont {R.}~\bibnamefont {Takahashi}},\ }\bibfield  {title} {\bibinfo {title} {{Probing Dark Low-mass Halos and Primordial Black Holes with Frequency-dependent Gravitational Lensing Dispersions of Gravitational Waves}},\ }\href {https://doi.org/10.3847/1538-4357/abafab} {\bibfield  {journal} {\bibinfo  {journal} {Astrophys. J.}\ }\textbf {\bibinfo {volume} {901}},\ \bibinfo {pages} {58} (\bibinfo {year} {2020})},\ \Eprint {https://arxiv.org/abs/2007.01936} {arXiv:2007.01936 [astro-ph.CO]} \BibitemShut {NoStop}%
\bibitem [{\citenamefont {Choi}\ \emph {et~al.}(2021)\citenamefont {Choi}, \citenamefont {Park},\ and\ \citenamefont {Jung}}]{Choi:2021bkx}%
  \BibitemOpen
  \bibfield  {author} {\bibinfo {author} {\bibfnamefont {H.~G.}\ \bibnamefont {Choi}}, \bibinfo {author} {\bibfnamefont {C.}~\bibnamefont {Park}},\ and\ \bibinfo {author} {\bibfnamefont {S.}~\bibnamefont {Jung}},\ }\bibfield  {title} {\bibinfo {title} {{Small-scale shear: Peeling off diffuse subhalos with gravitational waves}},\ }\href {https://doi.org/10.1103/PhysRevD.104.063001} {\bibfield  {journal} {\bibinfo  {journal} {Phys. Rev. D}\ }\textbf {\bibinfo {volume} {104}},\ \bibinfo {pages} {063001} (\bibinfo {year} {2021})},\ \Eprint {https://arxiv.org/abs/2103.08618} {arXiv:2103.08618 [astro-ph.CO]} \BibitemShut {NoStop}%
\bibitem [{\citenamefont {Tambalo}\ \emph {et~al.}(2022)\citenamefont {Tambalo}, \citenamefont {Zumalac{\'a}rregui}, \citenamefont {Dai},\ and\ \citenamefont {Cheung}}]{tambalo2022gravitational}%
  \BibitemOpen
  \bibfield  {author} {\bibinfo {author} {\bibfnamefont {G.}~\bibnamefont {Tambalo}}, \bibinfo {author} {\bibfnamefont {M.}~\bibnamefont {Zumalac{\'a}rregui}}, \bibinfo {author} {\bibfnamefont {L.}~\bibnamefont {Dai}},\ and\ \bibinfo {author} {\bibfnamefont {M.~H.-Y.}\ \bibnamefont {Cheung}},\ }\bibfield  {title} {\bibinfo {title} {Gravitational wave lensing as a probe of halo properties and dark matter},\ }\href@noop {} {\bibfield  {journal} {\bibinfo  {journal} {arXiv preprint arXiv:2212.11960}\ } (\bibinfo {year} {2022})}\BibitemShut {NoStop}%
\bibitem [{\citenamefont {Berezinsky}\ \emph {et~al.}(2013)\citenamefont {Berezinsky}, \citenamefont {Dokuchaev},\ and\ \citenamefont {Eroshenko}}]{Berezinsky:2013fxa}%
  \BibitemOpen
  \bibfield  {author} {\bibinfo {author} {\bibfnamefont {V.~S.}\ \bibnamefont {Berezinsky}}, \bibinfo {author} {\bibfnamefont {V.~I.}\ \bibnamefont {Dokuchaev}},\ and\ \bibinfo {author} {\bibfnamefont {Y.~N.}\ \bibnamefont {Eroshenko}},\ }\bibfield  {title} {\bibinfo {title} {{Formation and internal structure of superdense dark matter clumps and ultracompact minihaloes}},\ }\href {https://doi.org/10.1088/1475-7516/2013/11/059} {\bibfield  {journal} {\bibinfo  {journal} {JCAP}\ }\textbf {\bibinfo {volume} {11}},\ \bibinfo {pages} {059}},\ \Eprint {https://arxiv.org/abs/1308.6742} {arXiv:1308.6742 [astro-ph.CO]} \BibitemShut {NoStop}%
\bibitem [{\citenamefont {Boudaud}\ \emph {et~al.}(2021)\citenamefont {Boudaud}, \citenamefont {Lacroix}, \citenamefont {Stref}, \citenamefont {Lavalle},\ and\ \citenamefont {Salati}}]{Boudaud:2021irr}%
  \BibitemOpen
  \bibfield  {author} {\bibinfo {author} {\bibfnamefont {M.}~\bibnamefont {Boudaud}}, \bibinfo {author} {\bibfnamefont {T.}~\bibnamefont {Lacroix}}, \bibinfo {author} {\bibfnamefont {M.}~\bibnamefont {Stref}}, \bibinfo {author} {\bibfnamefont {J.}~\bibnamefont {Lavalle}},\ and\ \bibinfo {author} {\bibfnamefont {P.}~\bibnamefont {Salati}},\ }\bibfield  {title} {\bibinfo {title} {{In-depth analysis of the clustering of dark matter particles around primordial black holes. Part~I. Density profiles}},\ }\href {https://doi.org/10.1088/1475-7516/2021/08/053} {\bibfield  {journal} {\bibinfo  {journal} {JCAP}\ }\textbf {\bibinfo {volume} {08}},\ \bibinfo {pages} {053}},\ \Eprint {https://arxiv.org/abs/2106.07480} {arXiv:2106.07480 [astro-ph.CO]} \BibitemShut {NoStop}%
\bibitem [{\citenamefont {Deguchi}\ and\ \citenamefont {Watson}(1986)}]{deguchi1986wave}%
  \BibitemOpen
  \bibfield  {author} {\bibinfo {author} {\bibfnamefont {S.}~\bibnamefont {Deguchi}}\ and\ \bibinfo {author} {\bibfnamefont {W.~D.}\ \bibnamefont {Watson}},\ }\bibfield  {title} {\bibinfo {title} {Wave effects in gravitational lensing of electromagnetic radiation},\ }\href@noop {} {\bibfield  {journal} {\bibinfo  {journal} {Physical Review D}\ }\textbf {\bibinfo {volume} {34}},\ \bibinfo {pages} {1708} (\bibinfo {year} {1986})}\BibitemShut {NoStop}%
\bibitem [{\citenamefont {Nakamura}\ and\ \citenamefont {Deguchi}(1999)}]{nakamura1999wave}%
  \BibitemOpen
  \bibfield  {author} {\bibinfo {author} {\bibfnamefont {T.~T.}\ \bibnamefont {Nakamura}}\ and\ \bibinfo {author} {\bibfnamefont {S.}~\bibnamefont {Deguchi}},\ }\bibfield  {title} {\bibinfo {title} {Wave optics in gravitational lensing},\ }\href@noop {} {\bibfield  {journal} {\bibinfo  {journal} {Progress of Theoretical Physics Supplement}\ }\textbf {\bibinfo {volume} {133}},\ \bibinfo {pages} {137} (\bibinfo {year} {1999})}\BibitemShut {NoStop}%
\bibitem [{\citenamefont {Macquart}(2004)}]{Macquart:2004sh}%
  \BibitemOpen
  \bibfield  {author} {\bibinfo {author} {\bibfnamefont {J.-P.}\ \bibnamefont {Macquart}},\ }\bibfield  {title} {\bibinfo {title} {{Scattering of gravitational radiation: Second order moments of the wave amplitude}},\ }\href {https://doi.org/10.1051/0004-6361:20034512} {\bibfield  {journal} {\bibinfo  {journal} {Astron. Astrophys.}\ }\textbf {\bibinfo {volume} {422}},\ \bibinfo {pages} {761} (\bibinfo {year} {2004})},\ \Eprint {https://arxiv.org/abs/astro-ph/0402661} {arXiv:astro-ph/0402661} \BibitemShut {NoStop}%
\bibitem [{\citenamefont {Jung}\ and\ \citenamefont {Kim}(2023)}]{Jung:2022tzn}%
  \BibitemOpen
  \bibfield  {author} {\bibinfo {author} {\bibfnamefont {S.}~\bibnamefont {Jung}}\ and\ \bibinfo {author} {\bibfnamefont {S.}~\bibnamefont {Kim}},\ }\bibfield  {title} {\bibinfo {title} {{Solar diffraction of LIGO-band gravitational waves}},\ }\href {https://doi.org/10.1088/1475-7516/2023/07/042} {\bibfield  {journal} {\bibinfo  {journal} {JCAP}\ }\textbf {\bibinfo {volume} {07}},\ \bibinfo {pages} {042}},\ \Eprint {https://arxiv.org/abs/2210.02649} {arXiv:2210.02649 [astro-ph.CO]} \BibitemShut {NoStop}%
\bibitem [{\citenamefont {Takahashi}(2004)}]{takahashi2004quasi}%
  \BibitemOpen
  \bibfield  {author} {\bibinfo {author} {\bibfnamefont {R.}~\bibnamefont {Takahashi}},\ }\bibfield  {title} {\bibinfo {title} {Quasi-geometrical optics approximation in gravitational lensing},\ }\href@noop {} {\bibfield  {journal} {\bibinfo  {journal} {Astronomy \& Astrophysics}\ }\textbf {\bibinfo {volume} {423}},\ \bibinfo {pages} {787} (\bibinfo {year} {2004})}\BibitemShut {NoStop}%
\bibitem [{\citenamefont {Savastano}\ \emph {et~al.}(2023)\citenamefont {Savastano}, \citenamefont {Tambalo}, \citenamefont {Villarrubia-Rojo},\ and\ \citenamefont {Zumalacarregui}}]{savastano2023weakly}%
  \BibitemOpen
  \bibfield  {author} {\bibinfo {author} {\bibfnamefont {S.}~\bibnamefont {Savastano}}, \bibinfo {author} {\bibfnamefont {G.}~\bibnamefont {Tambalo}}, \bibinfo {author} {\bibfnamefont {H.}~\bibnamefont {Villarrubia-Rojo}},\ and\ \bibinfo {author} {\bibfnamefont {M.}~\bibnamefont {Zumalacarregui}},\ }\bibfield  {title} {\bibinfo {title} {Weakly lensed gravitational waves: Probing cosmic structures with wave-optics features},\ }\href@noop {} {\bibfield  {journal} {\bibinfo  {journal} {arXiv preprint arXiv:2306.05282}\ } (\bibinfo {year} {2023})}\BibitemShut {NoStop}%
\bibitem [{\citenamefont {Cutler}\ and\ \citenamefont {Flanagan}(1994)}]{cutler1994gravitational}%
  \BibitemOpen
  \bibfield  {author} {\bibinfo {author} {\bibfnamefont {C.}~\bibnamefont {Cutler}}\ and\ \bibinfo {author} {\bibfnamefont {E.~E.}\ \bibnamefont {Flanagan}},\ }\bibfield  {title} {\bibinfo {title} {Gravitational waves from merging compact binaries: How accurately can one extract the binary's parameters from the inspiral waveform?},\ }\href@noop {} {\bibfield  {journal} {\bibinfo  {journal} {Physical Review D}\ }\textbf {\bibinfo {volume} {49}},\ \bibinfo {pages} {2658} (\bibinfo {year} {1994})}\BibitemShut {NoStop}%
\bibitem [{\citenamefont {Abbott}\ \emph {et~al.}(2020{\natexlab{a}})\citenamefont {Abbott} \emph {et~al.}}]{LIGOScientific:2019hgc}%
  \BibitemOpen
  \bibfield  {author} {\bibinfo {author} {\bibfnamefont {B.~P.}\ \bibnamefont {Abbott}} \emph {et~al.} (\bibinfo {collaboration} {LIGO Scientific, Virgo}),\ }\bibfield  {title} {\bibinfo {title} {{A guide to LIGO\textendash{}Virgo detector noise and extraction of transient gravitational-wave signals}},\ }\href {https://doi.org/10.1088/1361-6382/ab685e} {\bibfield  {journal} {\bibinfo  {journal} {Class. Quant. Grav.}\ }\textbf {\bibinfo {volume} {37}},\ \bibinfo {pages} {055002} (\bibinfo {year} {2020}{\natexlab{a}})},\ \Eprint {https://arxiv.org/abs/1908.11170} {arXiv:1908.11170 [gr-qc]} \BibitemShut {NoStop}%
\bibitem [{\citenamefont {Romano}\ and\ \citenamefont {Cornish}(2017)}]{romano2017detection}%
  \BibitemOpen
  \bibfield  {author} {\bibinfo {author} {\bibfnamefont {J.~D.}\ \bibnamefont {Romano}}\ and\ \bibinfo {author} {\bibfnamefont {N.~J.}\ \bibnamefont {Cornish}},\ }\bibfield  {title} {\bibinfo {title} {Detection methods for stochastic gravitational-wave backgrounds: a unified treatment},\ }\href@noop {} {\bibfield  {journal} {\bibinfo  {journal} {Living reviews in relativity}\ }\textbf {\bibinfo {volume} {20}},\ \bibinfo {pages} {1} (\bibinfo {year} {2017})}\BibitemShut {NoStop}%
\bibitem [{\citenamefont {Talbot}\ and\ \citenamefont {Thrane}(2018)}]{talbot2018measuring}%
  \BibitemOpen
  \bibfield  {author} {\bibinfo {author} {\bibfnamefont {C.}~\bibnamefont {Talbot}}\ and\ \bibinfo {author} {\bibfnamefont {E.}~\bibnamefont {Thrane}},\ }\bibfield  {title} {\bibinfo {title} {Measuring the binary black hole mass spectrum with an astrophysically motivated parameterization},\ }\href@noop {} {\bibfield  {journal} {\bibinfo  {journal} {The Astrophysical Journal}\ }\textbf {\bibinfo {volume} {856}},\ \bibinfo {pages} {173} (\bibinfo {year} {2018})}\BibitemShut {NoStop}%
\bibitem [{\citenamefont {Abbott}\ \emph {et~al.}(2019)\citenamefont {Abbott}, \citenamefont {Abbott}, \citenamefont {Abbott}, \citenamefont {Abraham}, \citenamefont {Acernese}, \citenamefont {Ackley}, \citenamefont {Adams}, \citenamefont {Adhikari}, \citenamefont {Adya}, \citenamefont {Affeldt} \emph {et~al.}}]{abbott2019binary}%
  \BibitemOpen
  \bibfield  {author} {\bibinfo {author} {\bibfnamefont {B.}~\bibnamefont {Abbott}}, \bibinfo {author} {\bibfnamefont {R.}~\bibnamefont {Abbott}}, \bibinfo {author} {\bibfnamefont {T.}~\bibnamefont {Abbott}}, \bibinfo {author} {\bibfnamefont {S.}~\bibnamefont {Abraham}}, \bibinfo {author} {\bibfnamefont {F.}~\bibnamefont {Acernese}}, \bibinfo {author} {\bibfnamefont {K.}~\bibnamefont {Ackley}}, \bibinfo {author} {\bibfnamefont {C.}~\bibnamefont {Adams}}, \bibinfo {author} {\bibfnamefont {R.~X.}\ \bibnamefont {Adhikari}}, \bibinfo {author} {\bibfnamefont {V.}~\bibnamefont {Adya}}, \bibinfo {author} {\bibfnamefont {C.}~\bibnamefont {Affeldt}}, \emph {et~al.},\ }\bibfield  {title} {\bibinfo {title} {Binary black hole population properties inferred from the first and second observing runs of advanced ligo and advanced virgo},\ }\href@noop {} {\bibfield  {journal} {\bibinfo  {journal} {The Astrophysical Journal Letters}\ }\textbf {\bibinfo {volume} {882}},\ \bibinfo {pages} {L24} (\bibinfo {year}
  {2019})}\BibitemShut {NoStop}%
\bibitem [{\citenamefont {Abbott}\ \emph {et~al.}(2021{\natexlab{b}})\citenamefont {Abbott}, \citenamefont {Abbott}, \citenamefont {Abraham}, \citenamefont {Acernese}, \citenamefont {Ackley}, \citenamefont {Adams}, \citenamefont {Adams}, \citenamefont {Adhikari}, \citenamefont {Adya}, \citenamefont {Affeldt} \emph {et~al.}}]{abbott2021population}%
  \BibitemOpen
  \bibfield  {author} {\bibinfo {author} {\bibfnamefont {R.}~\bibnamefont {Abbott}}, \bibinfo {author} {\bibfnamefont {T.}~\bibnamefont {Abbott}}, \bibinfo {author} {\bibfnamefont {S.}~\bibnamefont {Abraham}}, \bibinfo {author} {\bibfnamefont {F.}~\bibnamefont {Acernese}}, \bibinfo {author} {\bibfnamefont {K.}~\bibnamefont {Ackley}}, \bibinfo {author} {\bibfnamefont {A.}~\bibnamefont {Adams}}, \bibinfo {author} {\bibfnamefont {C.}~\bibnamefont {Adams}}, \bibinfo {author} {\bibfnamefont {R.}~\bibnamefont {Adhikari}}, \bibinfo {author} {\bibfnamefont {V.}~\bibnamefont {Adya}}, \bibinfo {author} {\bibfnamefont {C.}~\bibnamefont {Affeldt}}, \emph {et~al.},\ }\bibfield  {title} {\bibinfo {title} {Population properties of compact objects from the second ligo--virgo gravitational-wave transient catalog},\ }\href@noop {} {\bibfield  {journal} {\bibinfo  {journal} {The Astrophysical journal letters}\ }\textbf {\bibinfo {volume} {913}},\ \bibinfo {pages} {L7} (\bibinfo {year} {2021}{\natexlab{b}})}\BibitemShut
  {NoStop}%
\bibitem [{\citenamefont {Abbott}\ \emph {et~al.}(2023{\natexlab{b}})\citenamefont {Abbott} \emph {et~al.}}]{KAGRA:2021duu}%
  \BibitemOpen
  \bibfield  {author} {\bibinfo {author} {\bibfnamefont {R.}~\bibnamefont {Abbott}} \emph {et~al.} (\bibinfo {collaboration} {KAGRA, VIRGO, LIGO Scientific}),\ }\bibfield  {title} {\bibinfo {title} {{Population of Merging Compact Binaries Inferred Using Gravitational Waves through GWTC-3}},\ }\href {https://doi.org/10.1103/PhysRevX.13.011048} {\bibfield  {journal} {\bibinfo  {journal} {Phys. Rev. X}\ }\textbf {\bibinfo {volume} {13}},\ \bibinfo {pages} {011048} (\bibinfo {year} {2023}{\natexlab{b}})},\ \Eprint {https://arxiv.org/abs/2111.03634} {arXiv:2111.03634 [astro-ph.HE]} \BibitemShut {NoStop}%
\bibitem [{\citenamefont {Abbott}\ \emph {et~al.}(2020{\natexlab{b}})\citenamefont {Abbott}, \citenamefont {Abbott}, \citenamefont {Abbott}, \citenamefont {Abraham}, \citenamefont {Acernese}, \citenamefont {Ackley}, \citenamefont {Adams}, \citenamefont {Adya}, \citenamefont {Affeldt}, \citenamefont {Agathos} \emph {et~al.}}]{abbott2020prospects}%
  \BibitemOpen
  \bibfield  {author} {\bibinfo {author} {\bibfnamefont {B.~P.}\ \bibnamefont {Abbott}}, \bibinfo {author} {\bibfnamefont {R.}~\bibnamefont {Abbott}}, \bibinfo {author} {\bibfnamefont {T.}~\bibnamefont {Abbott}}, \bibinfo {author} {\bibfnamefont {S.}~\bibnamefont {Abraham}}, \bibinfo {author} {\bibfnamefont {F.}~\bibnamefont {Acernese}}, \bibinfo {author} {\bibfnamefont {K.}~\bibnamefont {Ackley}}, \bibinfo {author} {\bibfnamefont {C.}~\bibnamefont {Adams}}, \bibinfo {author} {\bibfnamefont {V.}~\bibnamefont {Adya}}, \bibinfo {author} {\bibfnamefont {C.}~\bibnamefont {Affeldt}}, \bibinfo {author} {\bibfnamefont {M.}~\bibnamefont {Agathos}}, \emph {et~al.},\ }\bibfield  {title} {\bibinfo {title} {Prospects for observing and localizing gravitational-wave transients with advanced ligo, advanced virgo and kagra},\ }\href@noop {} {\bibfield  {journal} {\bibinfo  {journal} {Living reviews in relativity}\ }\textbf {\bibinfo {volume} {23}},\ \bibinfo {pages} {1} (\bibinfo {year} {2020}{\natexlab{b}})}\BibitemShut
  {NoStop}%
\bibitem [{\citenamefont {Hild}\ \emph {et~al.}(2011)\citenamefont {Hild}, \citenamefont {Abernathy}, \citenamefont {Acernese}, \citenamefont {Amaro-Seoane}, \citenamefont {Andersson}, \citenamefont {Arun}, \citenamefont {Barone}, \citenamefont {Barr}, \citenamefont {Barsuglia}, \citenamefont {Beker} \emph {et~al.}}]{hild2011sensitivity}%
  \BibitemOpen
  \bibfield  {author} {\bibinfo {author} {\bibfnamefont {S.}~\bibnamefont {Hild}}, \bibinfo {author} {\bibfnamefont {M.}~\bibnamefont {Abernathy}}, \bibinfo {author} {\bibfnamefont {F.~e.}\ \bibnamefont {Acernese}}, \bibinfo {author} {\bibfnamefont {P.}~\bibnamefont {Amaro-Seoane}}, \bibinfo {author} {\bibfnamefont {N.}~\bibnamefont {Andersson}}, \bibinfo {author} {\bibfnamefont {K.}~\bibnamefont {Arun}}, \bibinfo {author} {\bibfnamefont {F.}~\bibnamefont {Barone}}, \bibinfo {author} {\bibfnamefont {B.}~\bibnamefont {Barr}}, \bibinfo {author} {\bibfnamefont {M.}~\bibnamefont {Barsuglia}}, \bibinfo {author} {\bibfnamefont {M.}~\bibnamefont {Beker}}, \emph {et~al.},\ }\bibfield  {title} {\bibinfo {title} {Sensitivity studies for third-generation gravitational wave observatories},\ }\href@noop {} {\bibfield  {journal} {\bibinfo  {journal} {Classical and Quantum gravity}\ }\textbf {\bibinfo {volume} {28}},\ \bibinfo {pages} {094013} (\bibinfo {year} {2011})}\BibitemShut {NoStop}%
\bibitem [{\citenamefont {Srivastava}\ \emph {et~al.}(2022)\citenamefont {Srivastava}, \citenamefont {Davis}, \citenamefont {Kuns}, \citenamefont {Landry}, \citenamefont {Ballmer}, \citenamefont {Evans}, \citenamefont {Hall}, \citenamefont {Read},\ and\ \citenamefont {Sathyaprakash}}]{srivastava2022science}%
  \BibitemOpen
  \bibfield  {author} {\bibinfo {author} {\bibfnamefont {V.}~\bibnamefont {Srivastava}}, \bibinfo {author} {\bibfnamefont {D.}~\bibnamefont {Davis}}, \bibinfo {author} {\bibfnamefont {K.}~\bibnamefont {Kuns}}, \bibinfo {author} {\bibfnamefont {P.}~\bibnamefont {Landry}}, \bibinfo {author} {\bibfnamefont {S.}~\bibnamefont {Ballmer}}, \bibinfo {author} {\bibfnamefont {M.}~\bibnamefont {Evans}}, \bibinfo {author} {\bibfnamefont {E.~D.}\ \bibnamefont {Hall}}, \bibinfo {author} {\bibfnamefont {J.}~\bibnamefont {Read}},\ and\ \bibinfo {author} {\bibfnamefont {B.}~\bibnamefont {Sathyaprakash}},\ }\bibfield  {title} {\bibinfo {title} {Science-driven tunable design of cosmic explorer detectors},\ }\href@noop {} {\bibfield  {journal} {\bibinfo  {journal} {The Astrophysical Journal}\ }\textbf {\bibinfo {volume} {931}},\ \bibinfo {pages} {22} (\bibinfo {year} {2022})}\BibitemShut {NoStop}%
\bibitem [{\citenamefont {Urrutia}\ \emph {et~al.}(2023)\citenamefont {Urrutia}, \citenamefont {Vaskonen},\ and\ \citenamefont {Veerm\"ae}}]{Urrutia:2023mtk}%
  \BibitemOpen
  \bibfield  {author} {\bibinfo {author} {\bibfnamefont {J.}~\bibnamefont {Urrutia}}, \bibinfo {author} {\bibfnamefont {V.}~\bibnamefont {Vaskonen}},\ and\ \bibinfo {author} {\bibfnamefont {H.}~\bibnamefont {Veerm\"ae}},\ }\bibfield  {title} {\bibinfo {title} {{Gravitational wave microlensing by dressed primordial black holes}},\ }\href {https://doi.org/10.1103/PhysRevD.108.023507} {\bibfield  {journal} {\bibinfo  {journal} {Phys. Rev. D}\ }\textbf {\bibinfo {volume} {108}},\ \bibinfo {pages} {023507} (\bibinfo {year} {2023})},\ \Eprint {https://arxiv.org/abs/2303.17601} {arXiv:2303.17601 [astro-ph.CO]} \BibitemShut {NoStop}%
\end{thebibliography}%

\clearpage
\onecolumngrid
\begin{center}
   \textbf{\large SUPPLEMENTAL MATERIAL \\[.1cm] Coexistence Test of Primordial Black Holes and Particle Dark Matter\\from Diffractive Lensing}\\[.2cm]
  \vspace{0.05in}
  {Han Gil Choi, Sunghoon Jung, Philip Lu, Volodymyr Takhistov}
\end{center}

\twocolumngrid
\setcounter{equation}{0}
\setcounter{figure}{0}
\setcounter{table}{0}
\setcounter{section}{0}
\setcounter{page}{1}
\makeatletter
\renewcommand{\theequation}{S\arabic{equation}}
\renewcommand{\thefigure}{S\arabic{figure}}
\renewcommand{\thetable}{S\arabic{table}}

\onecolumngrid

We provide additional details of GW lensing, distinguishability of dressed and bare PBHs as well as sensitivity analysis.  
\section{Distinguishability of Dressed and Bare PBHs} \label{app:distinguish}

\subsection{Fitting geometric optics regime}

In the geometric optics limit ($f\rightarrow \infty$) of GW lensing, the solution of the Fresnel-Kirchhoff integral can be approximated by the usual two images~\cite{nakamura1999wave,takahashi2003wave}
\begin{equation}
    F(f) \simeq |\mu_1|^{1/2}+ |\mu_2|^{1/2} e^{2\pi i f \Delta t -i\frac{\pi}{2}},
\end{equation}
where $\mu_1$ and $\mu_2$ are the magnifications of the images, and $\Delta t$ is the arrival time difference between the lensing images. In GW observation of binary BH sources, the constant amplification due to lensing and the luminosity distance to the binary BH source are degenerate unless the source redshift is obtained from other observations. Therefore, the lensing amplification is reduced to
\begin{equation}
    F(f)\propto  1+\mu_r^{1/2}  e^{2\pi i f \Delta t -i\frac{\pi}{2}}~,
\end{equation}
and the only observables are the image magnification ratio $\mu_r = |\mu_1/\mu_2|$ and $\Delta t$.

A point mass (PBH) lens always produces two lensing images. From its image properties, one can find that
\begin{equation}
    \mu_r = \frac{1-\left(\sqrt{y_s^2/4+1}+y_s/2\right)^{-1}}{1-\left(\sqrt{y_s^2/4+1}-y_s/2\right)^{-1}}\, ~,
\end{equation}
and
\begin{equation}
    \Delta t = 4 M_l\bigg[y_s\sqrt{y_s^2/4+1} +\ln\left(1+y_s^2/2 + y_s \sqrt{y_s^2/4+1}\right)\bigg]\, ,
\end{equation}
where $y_s = x_s/x_E$, and $M_l= M_\text{PBH}(1+z_l)$. Note that relationship between the pairs of numbers $(\mu_r,\, \Delta t)$ and $(y_s,\, M_l)$ is invertible, and one can find the inverse relation
\begin{equation}
\begin{split}
    y_s &= \sqrt{\mu_r^{1/2} +\mu_r^{-1/2} -2 }\\
    M_l &= \frac{\Delta t}{2\left(\sqrt{\mu_r+\mu_r^{-1}-2}-\ln \mu_r \right) }\, .
    \end{split}
\end{equation}
This simple relation implies that any lensing signal induced by the interference of two lensing images can be interpreted as point mass lensing. Hence, in this situation, there is no distinguishability and different lens profiles are degenerate, unless additional information is given.

\subsection{Strong diffraction versus weak diffraction}

In our work, the weak diffraction (WD) and beyond Geometric Optics (bGO) signatures enable us to discriminate dPBHs from bare PBHs and others. Non-locally determined by lens profile, they induce nontrivial frequency-dependent behaviors. 

\begin{figure}
\includegraphics[width=0.5\linewidth]{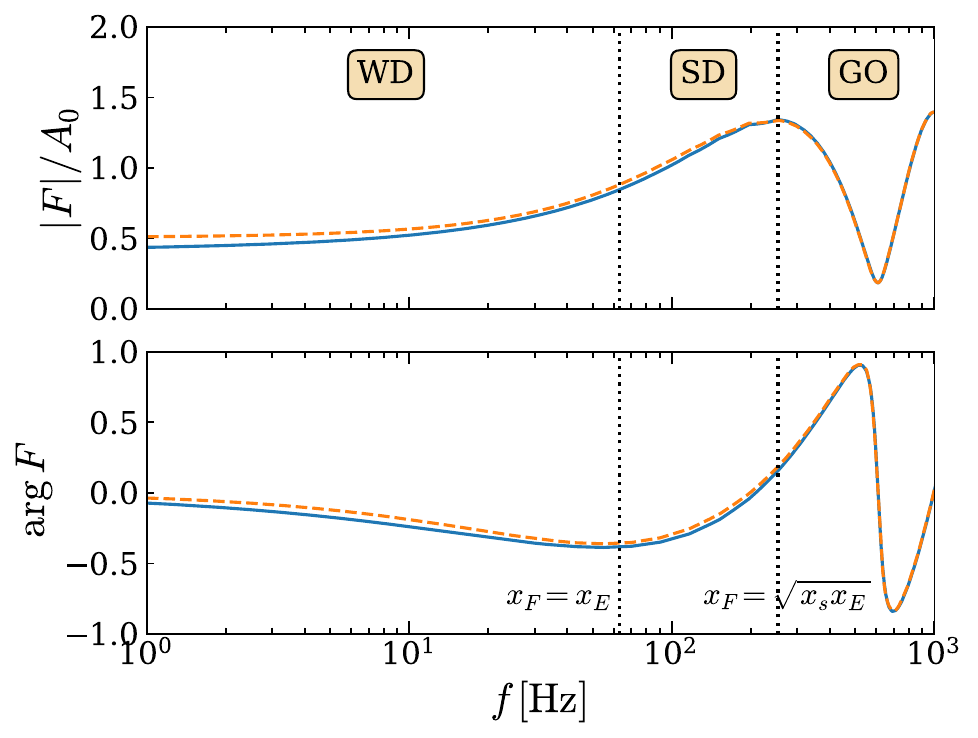}
\caption{\label{fig:SD}  Same as Fig.~2, but for the case with $x_s < x_E$, which induces strong diffraction for the intermediate regime $\sqrt{x_s x_E} \leq x_F \leq x_E$. This regime does not facilitate discrimination, once the GO regime is well fitted, but it still helps lensing detection due to the nontrivial frequency dependence over the unlensed $F/A_0=1$. }
\end{figure}

 Strong diffraction is also a non-local phenomenon, which occurs for the case with $x_s < x_E$. But this effect does not contribute to distinguishability since it is related to the blurring of the Einstein ring, where any symmetric strong lensing system has similar properties~\cite{Choi:2021bkx}. This is illustrated in \Fig{fig:SD}, which is produced with the same parameters as in Fig.~2 of the main text but in the case of $x_s < x_E$. Instead of bGO, strong diffraction occurs in the frequency range satisfying $\sqrt{x_s x_E}<x_F < x_E$. In this regime, $|F|\propto f^{1/2}$ for any symmetric lens profiles~\cite{Choi:2021bkx}. As a result, there is no distinction between dPBH lensing and PBH lensing (dashed) in this regime once data are matched at the geometric optics limit.

\subsection{Observed events with discrimination capability}

As a proxy of discrimination efficiency, we consider the fraction of observed GW lensing events that contain enough weak diffraction and bGO regimes in measured bands. Our analysis serves as a pilot study for a more dedicated analyses in the future.
More specifically, we estimate the discrimination fraction of detectable lensing events by the following two conditions
\begin{subequations}
\begin{equation}\label{eq:idc1}
x_F^\mathrm{max} \,>\, x_E,
\end{equation}
\begin{equation}\label{eq:idc2}
\rho_\mathrm{WD+bGO} \, \equiv\,  \sqrt{4\int_{f_\mathrm{min}}^{f_e}df\frac{|h_0(f)|^2}{S_n(f)}} \,>\, 10\, ,
\end{equation}
\end{subequations}
where $x_F^\mathrm{max}$ is the Fresnel length at the lowest sensitive frequency of the GW detector, and $f_e$ is the GW frequency at the $x_F=x_E$ instance. $x_E$ is the Einstein radius of dPBH. 

The former condition \Eq{eq:idc1} approximately corresponds to requiring that WD and bGO are included in the measurement band. Since the bGO effect outside the frequency range defined by \Eq{eq:idc1} is small but non-zero, it could provide some discrimination power; \Eq{eq:idc1} is a conservative requirement that can be further improved in dedicated analyses. The latter condition Eq.~\eqref{eq:idc2} approximately requires that the GW amplitude resolution in the weak diffraction plus bGO regime be better than $\sim10$\% level. This is approximately the scale of differences observed in Fig.~2, which we also expect to be improved with a dedicated study.

Within the region of the detectable lensing cross-section, we obtain the area of the region satisfying Eq.~\eqref{eq:idc1} and ~\eqref{eq:idc2}, which defines the discrimination cross-section, similarly to the lensing cross-section. Subsequently, we compute associated discrimination optical depths and discrimination rates. 
The discrimination rate divided by the lensing detection rate is the discrimination fraction of detectable lensing events.

Fig.~4 of the main text shows the 
the discrimination fraction of detectable lensing events for Cosmic Explorer (CE, solid), Einstein Telescope(ET, dashed), aLIGO A+(dotted), and aLIGO(dot-dashed). We find that the discrimination fraction sharply drops near $M_\mathrm{PBH}\sim \mathcal{O}(10-10^2)\, M_\odot$, where the Einstein radius of dPBH becomes comparable to the Fresnel lengths of GW spectrum. Above this mass range, lensing signals containing only geometric optics signatures are more frequent due to the large Einstein radius.

\begin{figure}
\includegraphics[width=0.45\linewidth]{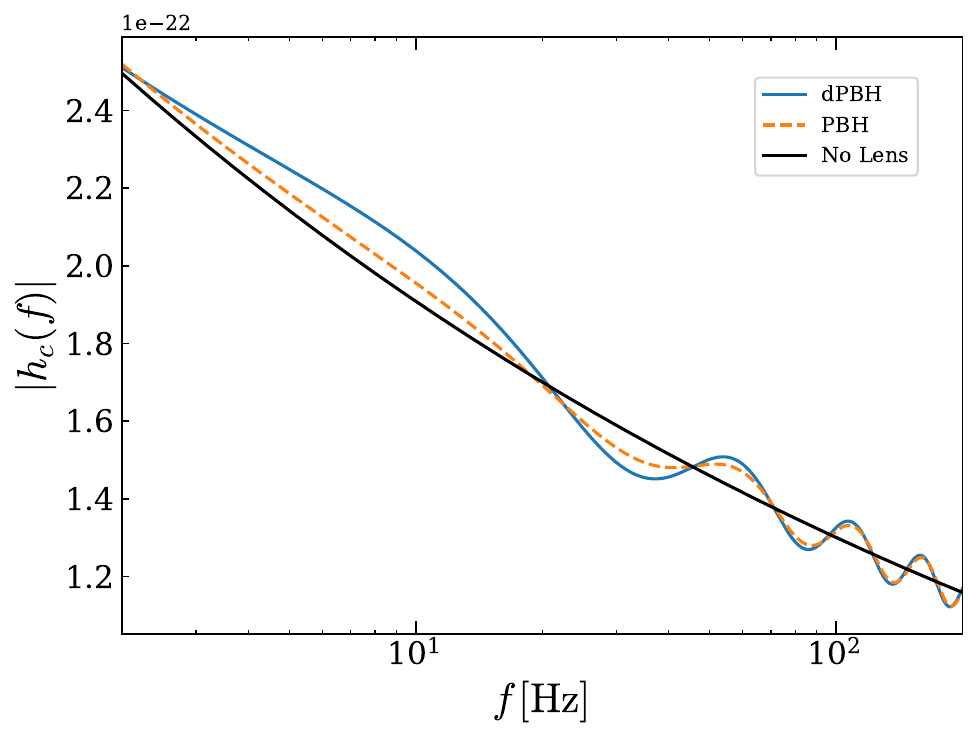} \includegraphics[width=0.45\linewidth]{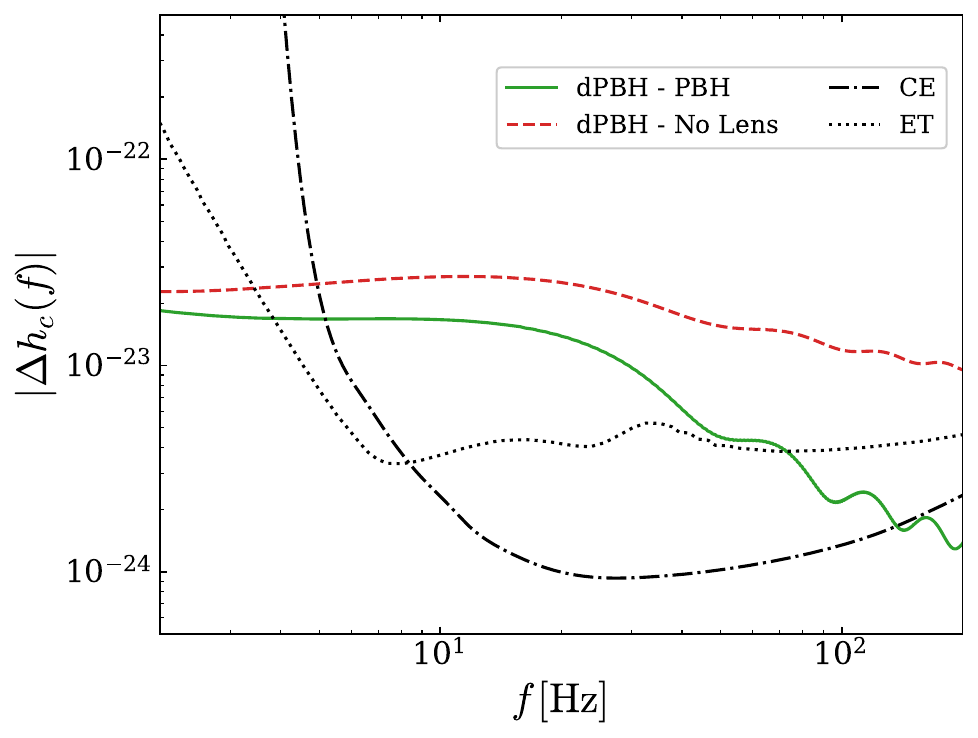}
\caption{\label{fig:strain}  
(Left panel) chirping GW spectrums assuming dPBH lensing (blue solid), bare PBH lensing (orange dashed), and no lensing (black solid). The characteristic strain $h_c(f)\equiv f h(f)$ is used. The GW source is $10 M_\odot$-$10 M_\odot$ binary BH at $z_s=0.4$. The lensing amplifications are given by $M_\text{PBH}=20 M_\odot$ and $z_l=0.17$; The lensing parameters are the same as those in Fig. 2 of the main text. (Right panel) GW strain difference $\Delta h_c \equiv |h^1_c - h^2_c|$ between the curves given in the left panel. The difference between the dPBH and PBH (green solid) is a signal of discrimination, while the difference between the dPBH and no lensing (red dashed) is a signal of lensing detection. We also present the noise amplitude $\sqrt{f S_n(f)}$ of CE (black dot-dashed) and ET (black dotted) for comparison. 
}
\end{figure}

Figure \ref{fig:strain} shows a typical example of the lensed GW strain that satisfies our discrimination criteria. We assume a chirping GW originated from a $10 M_\odot$-$10 M_\odot$ BH binary at $z_s=0.4$. The noticeable discrepancy between dPBH and PBH in the low-frequency region is due to diffractive lensing. Its magnitude is far above the noise amplitudes of CE and ET, hence it is measurable in those detectors. In the high-frequency region where lensing is described by GO, the discrepancy is buried under the detector noise and does not contribute to discrimination. Although the GO regime does not have discrimination power, it is a useful feature for lensing detection. \Fig{fig:strain} shows that the difference between dPBH lensing and no lensing is comparable at both high and low frequencies, equally contributing to lensing detection.

\section{Detection significance}  \label{app:significance}

\subsection{Likelihood ratio}

Our definition of lensing detection significance is based on the likelihood ratio 
\begin{equation}\label{eq:lambda}
    \Lambda \equiv \frac{p(d|\mathcal{H}_L)}{p(d|\mathcal{H}_0)}
\end{equation}
between the lensing detection hypothesis $\mathcal{H}_L$ and the no lensing hypothesis $\mathcal{H}_0$ given the strain data $d$. The probability $p(d|\mathcal{H}_0)$ and $p(d|\mathcal{H}_1)$ are the marginal likelihoods~\cite{romano2017detection}. The quantity Eq.~\eqref{eq:lambda} is also known as the Bayes factor. In principle, the marginal likelihoods should be computed by the integral of posterior distribution times prior distribution, $p(d|\mathcal{H})=\int d\bm{\theta} p(\bm{\theta}|d,\mathcal{H})p(\bm{\theta},\mathcal{H})$, where $\bm{\theta}$ is the model parameters given a hypothesis $\mathcal{H}$. However, the computation of the integral for many lensing data realizations requires prohibitively large computing resources. Instead, we use an approximate version of Eq.~\eqref{eq:lambda}~\cite{romano2017detection,dai2018detecting},
\begin{equation}
    \ln \Lambda \simeq \frac{1}{2}\left(\rho_{mL}^2 - \rho_{m0}^2 \right)\, ,
\end{equation}
where 
\begin{equation} \label{eq:mf}    
    \rho_{m}^2 \equiv \max \limits_{\bm{\theta}} \left[(d|d)-(d-h(\bm{\theta})|d-h(\bm{\theta}))\right]\\
\end{equation}
is the square of the matched filter Signal-to-Noise Ratio(SNR) with respect to waveform $h$. Here, the inner product $(
\cdot|\cdot)$ is defined as $(a|b) \equiv 4\text{Re}\int_0^{\infty} df\, a^*(f)b(f)/S_n(f)$ for given strain data $a$ and $b$ under stationary Gaussian detector noise with the noise spectral density $S_n(f)$. We denote $\rho_m$ by $\rho_{mL}$ and $\rho_{m0}$ for a lensed $h_L(\bm{\theta}_L)$ and unlensed $h_0(\bm{\theta_0})$ waveform, respectively. In these expressions, $\bm{\theta}_L$ and $\bm{\theta}_0$ are model parameters of $\mathcal{H}_L$ and $\mathcal{H}_0$, respectively.

Relying on the detector noise, the matched filter SNR and the likelihood ratio are stochastic quantities. Assuming the lensed waveform signal $h_L$ and the detector noise is small $d\sim h_L$, the expected value of the log-likelihood ratio is 
\begin{equation}\label{eq:lambdaexp}
    \langle \ln \Lambda \rangle \simeq \frac{1}{2}\left(\rho_L^2 - \rho_{uL}^2  \right)
     = \frac{1}{2}\min \limits_{\bm{\theta}_0} \left( h_L-h_0(\bm{\theta}_0)|h_L-h_0(\bm{\theta}_0) \right)\, ,
\end{equation}
where  $\rho_L^2 \equiv \langle\rho_{mL}^2 \rangle  \simeq (h_L|h_L)$, and $\rho_{uL}^2 \equiv \langle\rho_{m0}^2 \rangle  \simeq \max \limits_{\bm{\theta}_0} (h_0(\bm{\theta}_0)|2 h_L - h_0(\bm{\theta}_0))$. We use Eq.~\eqref{eq:lambdaexp} to define the lensing cross-section. 

\begin{figure}\label{fig:pf}
\includegraphics[width=0.5\linewidth]{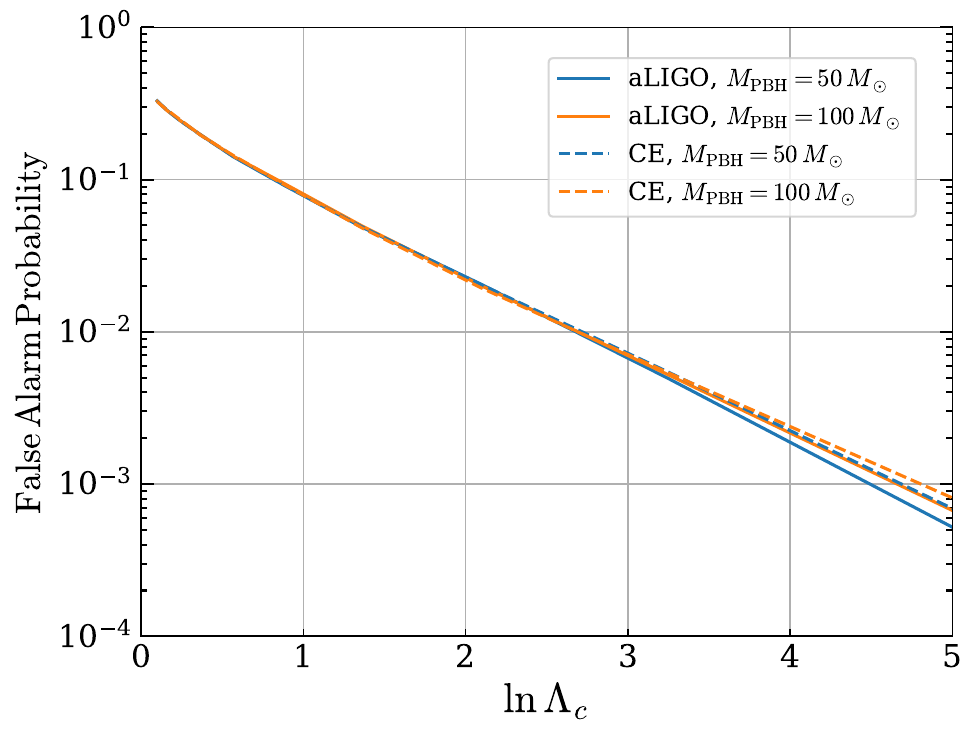}
\caption{The false-alarm probability for lensing detection criteria $\langle \ln \Lambda \rangle >\ln \Lambda_c$. In the cases of aLIGO (solid curves), $20\, M_\odot$ binary BH with $z_s=0.1$ is assumed, and we assume the dressed PBH lensing signals are falsely identified with $M_\mathrm{PBH}=50\, M_\odot$ (blue) and $M_\mathrm{PBH}=100\, M_\odot$ (orange). $z_l$ is set to give the maximum $d_\mathrm{eff}$ and $x_s$ is set to give $\langle \ln \Lambda \rangle=\ln \Lambda_c$. In the case of CE (dashed curves), only source redshift is changed to $z_s=1$. }
\end{figure}

In our work, we assume that a lensing signal $h_L$ which gives $\langle \ln \Lambda \rangle> \ln \Lambda_c$ is detectable. We should choose a large enough threshold $\ln \Lambda_c$ to avoid the situations when $\ln \Lambda > \ln \Lambda_c$ is satisfied even if the strain data does not contain any lensing signal. We use the False-alarm probability $P(\ln \Lambda >\ln \Lambda_c|\mathrm{no lensing})$ to justify our choice of $\ln \Lambda_c =3$. To compute the probability, we generate $10^5$ realizations of $d=n+h_0$ with a fixed $h_0$ and the stationary Gaussian detector noise $n$ and count the number of $\ln \Lambda>\ln \Lambda_c$ samples. In the computation of the $\rho_{mL}^2$, we choose $h_L(\bm{\theta}_L)$ that leads to $(\rho_L^2-\rho_{uL}^2)/2=\ln \Lambda_c$. To minimize the computation resources for the maximizations, we use the analytic method described in the next subsection. The False-alarm probability results for some combinations of lensing parameters are shown in \Fig{fig:pf}. We find that the False-alarm probabilities are not sensitive to lensing parameters or GW detectors and drop below $0.01$ for $\ln \Lambda_c= 3$ which is small enough for the discussions in our work.

\subsection{Optimization of the likelihood ratio}

We consider only three model parameters: constant amplitude factor $A$, constant phase shift $\theta_c$, and time shift $t_c$. We compute the matched filter SNR Eq. \eqref{eq:mf} by maximizing the following function:
\begin{equation}\label{eq:match}
    (d|d) - (d- A e^{i(\theta_c+w t_c)} h| d- A e^{i(\theta_c+w t_c)} h)
\end{equation}
for a given waveform $h$. Here we used $w=2\pi f$. The maximization of Eq.~\eqref{eq:match} with respect to $A$ and $\theta_c$ is given by 
\begin{equation}
    \frac{1}{(h|h)}[(d|e^{i w t_c}h)^2+(d|i e^{i w t_c} h)^2]\, .
\end{equation}
Maximization over $t_c$ requires numerical computation in general. However, when the expression is maximized at a small $t_c$, and we always inject the signal with $t_c=0$, one can obtain a leading order analytic expression for the $t_c$ maximization. The resulting analytic expression of $\rho_m^2$ is 
\begin{equation}
    \rho_m^2 \simeq \frac{1}{(h|h)}\bigg[ (d|h)^2 + (d|i h )^2+\frac{\left\{(d|h)(d|i w h) - (d|ih)(d|w h)\right\}^2}{(d|h)(d|w^2h)-(d|wh)^2+(d|ih)(d|iw^2h)-(d|iwh)^2}\bigg]~. 
\end{equation}
For a more detailed analysis, one can additionally consider binary BH intrinsic parameters such as total mass, mass ratio, black hole spins, etc, and lens intrinsic parameters such as lens mass, impact parameter, lens redshift, etc. For simplicity, we assume the parameters are not highly biased by the detector noise and are not correlated with each other. We leave a more detailed dedicated analysis for future works.

\subsection{Unlensed GW waveform model}

We use unlensed GW waveform in the Newtonian order to simplify our analysis. The waveform is given by \cite{cutler1994gravitational}
\begin{equation}
\begin{split}
    h_0(f) &= -  \sqrt{\frac{5}{24 }}\pi^{-2/3} A_p \frac{(\mathcal{M}_{z})^{5/6} }{d_L}f^{-7/6} e^{i \Psi(f)}\\
    \Psi(f) &=\theta_c + 2\pi f t_c -\frac{\pi}{4}+\frac{3}{128}(\pi \mathcal{M}_z  f)^{-5/3}
    \end{split}
\end{equation}
where we used $G=c=1$ units. Here, $d_L$ is the luminosity distance, and $\mathcal{M}_z = (m_1 m_2 )^{3/5}/(m_1+m_2)^{1/5}(1+z_s)$ is the detector-frame binary BH chirp mass defined by the component black hole masses $m_1$ and $m_2$. The polarization angle and detector orientation dependence are included in the $A_p$ factor, and we set $A_p=1$ for simplicity.

The upper cut-off of the GW frequency is set to the GW frequency at the inner-most-stable-circular-orbit frequency $f_\mathrm{isco}=(6^{3/2}\pi (m_1+m_2)(1+z_s))^{-1}$. The lower cut-off of the frequency follows the lower frequency bound of the GW detector sensitivity range. We adopt $5\, \mathrm{Hz}$ for Cosmic Explorer, $2\, \mathrm{Hz}$ for Einstein telescope, and $10\, \mathrm{Hz}$ for LIGO.

\section{Limit computation} \label{app:limitset}

We assume the occurrence of lensing events follows a Poisson distribution with the expected number of lensing detections $N_L = \int_0^{t_{\rm max}} dt \dot{N}_L$, taking $t_{\rm max} = 5$ years. Under the lensing detection hypothesis with $f_\mathrm{PBH}$ parameter, the probability of $k$ lensing detection is
\begin{equation}
    P(k|f_\mathrm{PBH}) = \frac{(f_\mathrm{PBH } \nu_L)^k}{k!}e^{-f_\mathrm{PBH }\nu_L}\, ,
\end{equation}
where $\nu_L = N_L(f_\mathrm{PBH}=1)$. Using Bayes' theorem, the posterior distribution of $f_\mathrm{PBH}$ is given by
\begin{equation}
    p(f_\mathrm{PBH}|k) =\frac{\pi(f_\mathrm{PBH}) P(k|f_\mathrm{PBH})}{P(k)} \, ,
\end{equation}
where $\pi(f_\mathrm{PBH})$ is the prior distribution of $f_\mathrm{PBH}$, and $P(k)\equiv \int d f_\mathrm{PBH} \pi(f_\mathrm{PBH}) P(k|f_\mathrm{PBH})$ is the marginalized likelihood.

The lensing detection rate $\dot{N}_L$ is given by
\begin{equation}
    \dot{N}_L = \int_{0}^{z_c}d V_c(z_s)\frac{R_0}{1+z_s} \langle \tau(z_s) \rangle\, ,
\end{equation}
where $dV_c(z_s)$ is the comoving volume element at $z_s$, and
\begin{equation}
        \langle \tau(z_s)\rangle = \int \limits_{\rho_0(M,\eta,z_s)>8}  dM d\eta  \, p_\mathrm{BBH}(M,\eta)  \tau(z_s;M,\eta)~ 
\end{equation}
is the expected lensing optical depth. Here $\rho_0(M,\eta,z_s)\equiv \sqrt{(h_0|h_0)}$ is the optimal SNR~\cite{LIGOScientific:2019hgc} of unlensed GW signal $h_0$ given by binary total mass $M=m_1+m_2$, mass ratio $\eta =(m_1 m_2)/M^2$, and redshift $z_s$.

The optical depth $\tau$ is approximately the detectable lensing probability for the given source, $1-e^{-\tau} \simeq \tau$, and it is obtained by
\begin{equation}
     \tau \,\simeq\, \int_0^{z_s} d\chi(z_l) ~n_\text{PBH} \, \sigma(z_l,z_s;M,\eta)~,
\end{equation}
with the comoving lensing cross-section,
\begin{equation} \label{eq:cs}
    \sigma \= (1+z_l)^2\, \int_{\langle \ln\Lambda \rangle >\ln \Lambda_c}d^2 \bm{x}_s~,
\end{equation}
measuring the comoving area of lens locations that can lead to detectable lensing. We used $\ln \Lambda_c=3$ for the detection threshold. The comoving number density of dPBH lenses, $n_{\rm PBH} = f_{\rm PBH} \Omega_{\rm DM} \rho_c/M_{\rm PBH}$, is assumed to be constant. $\rho_c$ is the critical density of the Friedmann universe.

Our choice of $p_\mathrm{BBH}(M,\eta)$ follows the Power law and Peak model~\cite{talbot2018measuring}, and we used $R_0 =28.3~\mathrm{Gpc}^{-3}\mathrm{yr}^{-1}$. We take into account binary BH merger sources up to redshift $z_c=10$ and assume their populations do not evolve with redshift. We assume the flat $\Lambda \mathrm{CDM}$ model with the Hubble parameter $H_0=67.74~\mathrm{km}~\mathrm{s}^{-1}~\mathrm{Mpc}^{-1}$, the matter density parameter $\Omega_m=0.3075$, and the dark matter density parameter $\Omega_\mathrm{DM}=0.2575$. 

We consider the null detection $k=0$ of lensing events to estimate the projected upper limit sensitivity of $f_\mathrm{PBH}$. We set a uniform prior on $f_\mathrm{PBH}$ with the range from 0 to 1. In this case, $P(0)=(1-e^{-\nu_L})/\nu_L$, and 
\begin{equation}
    p(f_\mathrm{PBH}|0) =\nu_L \frac{e^{-f_\mathrm{PBH}\nu_L}}{(1-e^{-\nu_L})}\, .
\end{equation}
Using the posterior, the 90\% upper limit of $f_\mathrm{PBH}$ is computed as
\begin{equation}
    f_\text{PBH}^{90\%} = -\frac{1}{\nu_L} \ln\left[1 - 0.9 (1 - e^{-\nu_L})\right]~.
\end{equation}

\end{document}